\documentclass[preprint,aps,prd,showpacs,letterpaper,nofootinbib]{revtex4}

\usepackage{graphicx}
\usepackage{amsmath,amsfonts,amssymb}

\def\msbar{{\overline{\rm MS}}}
\def\Bec{Be\'cirevi\'c}
\newcommand*{\chpt}{\raise0.4ex\hbox{$\chi$}PT}
\newcommand*{\schpt}{S\raise0.4ex\hbox{$\chi$}PT}
\newcommand*{\ie}{\textit{i.e.},\ }
\newcommand*{\eg}{\textit{e.g.},\ }

\newcommand*{\via}{\textit{via}}
\newcommand*{\et}{\textit{et al.}}

\newcommand*{\npb}[1]{Nucl.\ Phys.\ \textbf{B#1}}
\providecommand*{\prd}[1]{Phys.\ Rev.\ \textbf{D#1}}
\renewcommand*{\prd}[1]{Phys.\ Rev.\ \textbf{D#1}}

\newcommand*{\npbps}[1]{Nucl.\ Phys.\ \textbf{B}
        (Proc.\ Suppl.) \textbf{#1}}

\newcommand*{\tsukuba}{\npbps{129-130C} (2004)}
\def\fermilabtwo{Nucl.\ Phys.\ {\bf B} (Proc.\ Suppl.) {\bf 140} (2005)}
\def\tsukubathree{Nucl.\ Phys.\ {\bf B} (Proc.\ Suppl.) {\bf 129-130} (2004)}

\newcommand*{\MeV}{{\rm Me\!V}}

\newcommand{\opopo}{\ensuremath{1\!+\!1\!+\!1}}

\newcommand*{\Tr}{\ensuremath{\operatorname{Tr}}}

\newcommand{\trD}{\ensuremath{\textrm{tr}_{\textrm{\tiny \it D}}}}

\newcommand{\negcdot}{\negmedspace\cdot\negmedspace}
\newcommand{\vp}{\ensuremath{v\negcdot p}}
\newcommand{\vq}{\ensuremath{v\negcdot q}}
\newcommand{\vk}{\ensuremath{v\negcdot k}}

\newcommand{\Delstar}{\ensuremath{\Delta^{\raise0.18ex\hbox{${\scriptstyle *}$}}}}
\def\gtwid{{\,\raise.35ex\hbox{$>$\kern-.75em\lower1ex\hbox{$\sim$}}\,}}
\def\ltwid{{\,\raise.35ex\hbox{$<$\kern-.75em\lower1ex\hbox{$\sim$}}\,}}
\def\leftvec{{\raise1.5ex\hbox{$\leftarrow$}\kern-1.00em}}
\def\rightvec{{\raise1.5ex\hbox{$\rightarrow$}\kern-1.00em}}
\def\half{{\scriptstyle \raise.2ex\hbox{${1\over2}$}}}
\def\threehalves{{\scriptstyle \raise.15ex\hbox{${3\over2}$}}}
\def\third{{\scriptstyle \raise.15ex\hbox{${1\over3}$}}}
\def\third{{\scriptstyle \raise.15ex\hbox{${1\over3}$}}}
\def\twothirds{{\scriptstyle \raise.15ex\hbox{${2\over3}$}}}
\def\fourth{{\scriptstyle \raise.15ex\hbox{${1\over4}$}}}

\newcommand{\vslash}{\ensuremath{v\!\!\! /}}

\newcommand{\cA}{\ensuremath{\mathcal{A}}}

\newcommand{\cD}{\ensuremath{\mathcal{D}}}

\newcommand{\cF}{\ensuremath{\mathcal{F}}}

\newcommand{\cI}{\ensuremath{\mathcal{I}}}
\newcommand{\cJ}{\ensuremath{\mathcal{J}}}

\newcommand{\cL}{\ensuremath{\mathcal{L}}}
\newcommand{\cM}{\ensuremath{\mathcal{M}}}

\newcommand{\cO}{\ensuremath{\mathcal{O}}}

\newcommand{\cV}{\ensuremath{\mathcal{V}}}

\newcommand{\eqn}[1]{\label{eq:#1}}
\newcommand{\eq}[1]{Eq.~\eqref{eq:#1}}
\newcommand{\Eq}[1]{Equation \eqref{eq:#1}}
\newcommand{\eqs}[2]{Eqs.~\eqref{eq:#1} and \eqref{eq:#2}}

\newcommand{\eqsthru}[2] {Eqs.~\eqref{eq:#1} through \eqref{eq:#2}}
\newcommand{\eqsor}[2]{Eq.~\eqref{eq:#1} or \eqref{eq:#2}}
\newcommand{\eqsthree}[3]{Eqs.~\eqref{eq:#1}, \eqref{eq:#2} and \eqref{eq:#3}}

\newcommand{\eqsfour}[4]{Eqs.~\eqref{eq:#1}, \eqref{eq:#2}, \eqref{eq:#3},
and \eqref{eq:#4}}

\def\figref#1{Fig.~\ref{fig:#1}}
\def\figsref#1#2{Figs.~\ref{fig:#1} and \ref{fig:#2}}
\def\Figref#1{Figure~\ref{fig:#1}}

\def\secref#1{Sec.~\ref{sec:#1}}

\def\Secref#1{Section~\ref{sec:#1}}
\def\tabref#1{Table~\ref{tab:#1}}

\begin{document}

\title{Heavy-Light Semileptonic Decays in Staggered Chiral Perturbation Theory}
\author{C.~Aubin}
\affiliation{Physics Department, Columbia University, New York, NY 10027}
\author{C.~Bernard}
\affiliation{Department of Physics, Washington University, St.\ Louis, MO 63130}
\begin{abstract}
We calculate the form factors for the semileptonic decays of heavy-light pseudoscalar mesons
in partially quenched staggered chiral perturbation theory (\schpt), working to 
leading order in $1/m_Q$, where $m_Q$ is the heavy quark mass. We take the light meson
in the final state to be a pseudoscalar corresponding to the exact chiral symmetry
of staggered quarks.
The treatment assumes the validity of the standard prescription for representing the
staggered ``fourth root trick'' within \schpt\ by insertions of factors of $1/4$ for each sea quark loop.
Our calculation is based on an existing  
partially quenched continuum chiral perturbation theory calculation with degenerate sea quarks 
by \Bec, Prelovsek and Zupan, which we generalize to the staggered (and non-degenerate) case. 
As a by-product, we obtain the continuum partially quenched results with non-degenerate sea quarks.
We analyze the effects of non-leading chiral terms, and find a relation among the 
coefficients governing the analytic valence mass dependence at this order.
Our results are useful in analyzing lattice computations of form factors 
$B\to\pi$ and $D\to K$ when the light quarks are simulated with the staggered action.
\end{abstract}
\pacs{12.39.Fe,12.39.Hg, 11.30.Rd, 12.38.Gc}

\preprint{CU-TP-1177}

\maketitle

\section{Introduction}\label{sec:intro}

Extraction of the CKM matrix elements $|V_{ub}|$ and $|V_{cs}|$ from the experimentally measured 
semileptonic decay rates for 
$B\to \pi \ell\nu$  and $D\to K \ell\nu$ requires reliable theoretical calculations of the corresponding 
hadronic matrix elements.
Recently, there has been significant progress in computing these matrix elements
on the lattice, with good control of the systematic uncertainties \cite{ONOGI-lat06,OKAMOTO-lat05,WINGATE-REVIEW,KRONFELD-REVIEW}. 
Since computation time increases as a high power of the  inverse quark mass, 
the light ($u,d$) quark masses used in the simulations are heavier than in nature, and a
chiral extrapolation is necessary to obtain physical results.  To keep systematic errors small,
the simulated $u,d$ masses should be well into the chiral regime, giving pion masses $\sim\! 300\,\MeV$ or lighter.
Such masses in lattice calculations of leptonic and semileptonic heavy-light decays
are accessible with staggered quarks \cite{WINGATE,WINGATE_2,WINGATE_3,SHIGEMITSU,Aubin:2004ej,Aubin:2005ar}. 
The trade-off for this benefit is the fact that staggered quarks 
do not fully remove the species doubling that occurs for lattice fermions; for every flavor of lattice quark, 
there are four ``tastes,'' which are related in the continuum by an $SU(4)$ symmetry (or an
$SU(4)_L\times SU(4)_R$ symmetry in the massless case). 
The taste symmetry is broken at non-zero lattice spacing $a$ by terms of order $a^2$.

The breaking of taste symmetry  on the lattice implies that one must take into account taste-violations
in the chiral extrapolations, leading to a  
joint extrapolation in both the quark masses and the lattice spacing. 
Staggered chiral perturbation theory (\schpt) \cite{LEE_SHARPE,CHIRAL_FSB,SCHPT} allows us
to make such extrapolations systematic.
For quantities with heavy quarks, one must also incorporate Heavy Quark Effective Theory (HQET) 
\cite{Burdman:1992gh,Grinstein-et,Goity:1992tp,BOYD,MAN_WISE} into \schpt. This has been done in 
Ref.~\cite{HL_SCHPT}, and then applied to leptonic heavy-light decays. Here, we extend
the analysis of Ref.~\cite{HL_SCHPT} to the semileptonic case.

In addition to the practical implications of taste symmetry violations for chiral extrapolations,
the violations lead to a potentially more serious theoretical concern.
Simulations such as Refs.~\cite{WINGATE,WINGATE_2,WINGATE_3,SHIGEMITSU,Aubin:2004ej,Aubin:2005ar,FPI04} 
take the fourth root of the staggered quark determinant \cite{Marinari:1981qf} in an attempt to obtain a single
taste per quark flavor in the continuum limit. 
Were the taste symmetry exact at finite lattice spacing, the fourth root prescription would 
obviously accomplish the desired goal, since it would be equivalent to using a local
Dirac operator obtained by projecting the staggered operator onto a single-taste subspace.
Because the taste symmetry is broken, however, the fourth root is necessarily a nonlocal operation
at non-zero lattice spacing \cite{BGS}.
The question of whether the rooted theory is in the correct universality class therefore becomes
nontrivial. Nevertheless, there are strong 
theoretical arguments \cite{SHAMIR,CB-NF4,BGS,SHARPE-TALK,BGS-LAT06} in the interacting
theory, as well as free-theory and 
numerical evidence \cite{FOURTH-ROOT-NUMERICAL-AND-FREE}
that the fourth-root trick is valid, \ie that it produces QCD in the continuum limit.

The current paper does not actually need to assume that the rooting procedure itself
is valid.\footnote{Of course, were the fourth root trick to prove invalid, the {\it motivation}
for the current work would be lost.} Instead,  like previous \schpt\ calculations for the rooted
theory \cite{CHIRAL_FSB,SCHPT,HL_SCHPT,Laiho:2005np},
it requires a narrower assumption: that the rooting can be represented at the chiral level
by multiplying each sea quark loop by a factor of  $1/4$.  This can be accomplished
by a quark flow analysis  \cite{QUARK-FLOW}, or, more systematically,
by use of the replica trick \cite{REPLICA}.  In Ref.~\cite{CB-NF4}, it was shown
that the correctness of this  representation of the fourth root in \schpt\ 
follows in turn from certain --- in our opinion, rather plausible --- assumptions. 
As such, 
we assume here that this representation  is valid.
Fitting lattice quantities to \schpt\ formulae (as 
in Refs.~\cite{FPI04,Aubin:2005ar}) provides an additional empirical test of the validity of this 
representation.

The main purpose of the current paper is to find \schpt\ expressions for
the form factors of the semileptonic decay $B\to P\ell \nu$, 
where $P$ is some light pseudoscalar meson, 
which we will refer to generically as a ``pion.''
We consider first the partially quenched case, and obtain the full QCD results afterward by taking the 
limit where valence masses equal the sea masses.
The $B$ is a heavy-light meson made up of a $b$ heavy quark and a valence light quark spectator
of flavor $x$;
we use the notation $B_x$ when confusion as to the identity of the light spectator could arise.
The $P$ meson (more precisely $P_{xy}$) is composed of two light valence quarks, of flavor $x$ and $y$. 
For simplicity we consider only the case where the outgoing pion is (flavor) charged; in other
words $x\not=y$.
The flavor structure of the weak current responsible for the decay is $\bar y \gamma_\mu b$.

In our calculation, we take the heavy quark mass $m_Q$ to be large compared to $\Lambda_{\rm QCD}$ and 
work to leading order in the $1/m_Q$ expansion. 
Our analysis also applies when  the heavy quark is a $c$ (\ie to 
$D$ mesons), 
but we use $B$ to denote the heavy-light meson to stress the 
fact that only lowest order terms HQET are kept. 
For $D$ mesons, of course,  the higher order terms omitted here
would be more important than for $B$ mesons.

Discretization errors
coming from the heavy quark are not included in the current calculations.  
We assume that such errors will be estimated independently, using HQET as the
effective-theory description of the lattice heavy
quark \cite{Kronfeld:2000ck}. It is expected that the errors
from staggered quark taste-violations, which are considered here, are significantly more important
at most currently accessible lattice spacings than the heavy-quark errors \cite{KRONFELD-REVIEW}.
However, since taste-violations decrease rapidly\footnote{Taste violations
with improved staggered fermions go like $\alpha_S^2 a^2$.  See Fig.~1 in Ref.~\cite{MILC-LAT06}
for a test of this relation.}
when the lattice 
spacing is reduced, this may change  in the not too distant future.
In any case, the precise quantification of the total discretization error will 
always require simulation at several lattice spacings.

An additional practical constraint on the current calculation is that $am_Q$ 
must not be too large compared to unity.  When $am_Q\gg 1$, the effects of
the heavy quark doublers would need to be included in the chiral theory, and the
analysis would become prohibitively complicated.
A detailed discussion of this and other issues involved in incorporating heavy 
quarks into \schpt\ appears in Ref.~\cite{HL_SCHPT}. 

The calculations of interest here have been performed
in continuum partially quenched chiral perturbation theory (PQ\chpt) by 
\Bec, Prelovsek and Zupan \cite{BECIREVIC} for $N_{\rm sea}$ degenerate sea quarks. 
In this paper we show how one can generalize the PQ\chpt\ formulae to the corresponding \schpt\ formulae,
thereby avoiding the necessity of recomputing all the diagrams from scratch.

Some results from the current work, as well as a brief discussion of
how to generalize PQ\chpt\ to \schpt, appear in Ref.~\cite{Aubin:2004xd}. 
In addition, 
our results have already been used in chiral fits to lattice data 
in Refs.~\cite{Aubin:2004ej,SHIGEMITSU}. 
A related calculation for the $B\to D^*$ and $B\to D$ semileptonic form factors has been
presented by Laiho and Van de Water \cite{Laiho:2005np}.

The outline of this paper is as follows: 
We first include a brief description of heavy-light \schpt\ in Sec.~\ref{sec:hqetschpt}. 
In Sec.~\ref{sec:PQCHPTtoSCHPT}, we discuss the procedure for generalizing PQ\chpt\ to \schpt, 
using the heavy-light form factors as examples, 
although the procedure can be used for many other quantities in \schpt. 
Using this procedure and starting from Ref.~\cite{BECIREVIC}, 
we write down, in \secref{form}, the one-loop \schpt\ results 
for the semileptonic form factors. 
The partially quenched staggered case with non-degenerate sea quarks, as well as its
continuum limit, is presented in \secref{PQschpt}. 
In that section, we also
discuss a  method for treating --- in a way that appears to be practical 
for fitting lattice data --- some spurious singularities which arise in the 
calculations.  \Secref{fullQCD} considers full-QCD special cases of the
results from \secref{PQschpt}; while \secref{analytic} discusses the analytic
contributions to the form factors at this order.
In Sec.~\ref{sec:FV} we add in the effects of a finite spatial lattice volume. 
Sec.~\ref{sec:conc} presents our conclusions. 
We include three appendices: Appendix~\ref{app:rules} gives expressions for the \schpt\ propagators
and vertices, as well as the corresponding continuum versions.
Appendix~\ref{app:int} lists the integrals used in the form factor calculations; 
while Appendix~\ref{app:wf_ren} collects necessary wavefunction renormalization factors that were 
calculated in Refs.~\cite{SCHPT,HL_SCHPT}.

\section{Heavy-Light Staggered Chiral Perturbation 
Theory}\label{sec:hqetschpt}

References~\cite{Burdman:1992gh,Grinstein-et,Goity:1992tp,BOYD,MAN_WISE} show how
to incorporate heavy-light mesons 
into continuum \chpt; the extension to
\schpt\ appears in Ref.~\cite{HL_SCHPT}. 
Here we review the key features needed for our calculations.

The heavy-light vector ($B_{\mu a}^*$)
and pseudoscalar ($B_a$) mesons are combined in the field
\begin{equation}
  H_a = \frac{1 + \vslash}{2}\left[ \gamma_\mu B^{*}_{\mu a}
    + i \gamma_5 B_{a}\right]\ ,
\end{equation}
which destroys a meson.  Here
$v$ is the meson velocity, and $a$ is the ``flavor-taste''
index of the light quark in the meson. 
For $n$ flavors of light quarks, $a$ can take on $4n$ values. 
Later, we will write $a$ as separate flavor ($x$) and taste ($\alpha$) indices, 
$a\to (x,\alpha)$, 
and ultimately drop the taste index, since the quantities we calculate will have trivial
dependence on the light quark taste.
The conjugate field $\overline{H}_a$ creates mesons:
\begin{equation}
  \overline{H}_a \equiv \gamma_0 H^{\dagger}_a\gamma_0 =
  \left[ \gamma_\mu B^{\dagger *}_{\mu a}
    + i \gamma_5 B^{\dagger}_{a}\right]\frac{1 + \vslash}{2}\ .
\end{equation} 
As mentioned in the introduction, 
we use $B$ to denote generic 
heavy-light mesons 
to emphasize that we are working to leading order in $1/m_Q$.

Under $SU(2)$ heavy-quark spin symmetry, 
the heavy-light field
transforms as
\begin{eqnarray}
  H &\to & S H\ , \nonumber\\
  \overline{H} &\to & \overline{H}S^{\dagger}\ ,
\end{eqnarray}
with $S\in SU(2)$, 
while under
the $SU(4n)_L\times SU(4n)_R$ chiral symmetry,
\begin{eqnarray}
  H &\to & H \mathbb{U}^{\dagger}\ ,\nonumber\\
  \overline{H} &\to & \mathbb{U}\overline{H}\ ,
\end{eqnarray}
with $\mathbb{U}\in SU(4n)$ defined below. 
We keep the flavor and taste indices implicit here. 

The light mesons are combined in a Hermitian field $\Phi(x)$.
For $n$ staggered flavors, $\Phi$ is a $4n \times 4n$
matrix given by:
\begin{eqnarray}\label{eq:Phi}
  \Phi = \left( \begin{array}{cccc}
      U  & \pi^+ & K^+ & \cdots \\*
      \pi^- & D & K^0  & \cdots \\*
      K^-  & \bar{K^0}  & S  & \cdots \\*
      \vdots & \vdots & \vdots & \ddots \end{array} \right)\ .
\end{eqnarray}
We show the $n=3$ portion of $\Phi$ explicitly. 
Each entry in \eq{Phi} is a  $4\!\times\!4$ matrix, written
in terms of the 16 Hermitian taste
generators $T_\Xi$ as, for example,
$U = \sum_{\Xi=1}^{16} U_\Xi T_\Xi$. 
The component fields of the flavor-neutral
elements ($U_\Xi$, $D_\Xi$, \dots) are real; the other (flavor-charged)
fields ($\pi^+_\Xi$,  $K^0_\Xi$, \dots) are complex. 
 The $T_\Xi$ are
\begin{equation}\label{eq:T_Xi}
  T_\Xi = \{ \xi_5, 
i\xi_{\mu 5},
  i\xi_{\mu\nu} (\mu<\nu), \xi_{\mu},
  \xi_I\}\ ,
\end{equation}
with $\xi_\mu$ the taste matrices corresponding to the Dirac gamma matrices,
and $\xi_I \equiv I$ the $4\times 4$ identity matrix.
We define $\xi_{\mu5}\equiv \xi_{\mu}\xi_5$, and 
$\xi_{\mu\nu}\equiv (1/2)[\xi_{\mu},\xi_{\nu}]$. 

 The
mass matrix is the $4n\times 4n$ matrix
\begin{eqnarray}
  \cM = \left( \begin{array}{cccc}
      m_u I  & 0 &0  & \cdots \\*
      0  & m_d I & 0  & \cdots \\*
      0  & 0  & m_s I  & \cdots\\*
      \vdots & \vdots & \vdots & \ddots \end{array} \right),
\end{eqnarray}
where 
the portion shown is again for the $n=3$ case.

  From $\Phi$ one constructs the 
unitary chiral field $\Sigma = \exp [i\Phi/f]$,
with $f$ the tree-level pion decay constant. In our normalization, $f \sim f_\pi \cong 131\ \MeV$.
Terms involving the heavy-lights are conveniently written using 
use $\sigma \equiv \sqrt{\Sigma} = \exp[ i\Phi / 2f ]$. 
These fields transform trivially under
the $SU(2)$ spin symmetry, 
while under $SU(4n)_L\times SU(4n)_R$ we have
\begin{eqnarray}
  \Sigma \to  L\Sigma R^{\dagger}\,,\qquad&&\qquad
  \Sigma^\dagger \to  R\Sigma^\dagger L^{\dagger}\,,\\*
  \sigma \to  L\sigma \mathbb{U}^{\dagger} = \mathbb{U} \sigma R^{\dagger}\,, \qquad&&\qquad
  \sigma^\dagger \to R \sigma^\dagger \mathbb{U}^{\dagger} = \mathbb{U} \sigma^\dagger L^{\dagger}\,, 
  \label{eq:Udef}
\end{eqnarray}
with global transformations $L\in SU(4n)_L$ and $R\in SU(4n)_R$.
The transformation $\mathbb{U}$, defined by \eq{Udef}, is
is a function of $\Phi$ and therefore 
of the coordinates.

It is convenient to define objects involving the $\sigma$ field that transform 
only with $\mathbb{U}$ and $\mathbb{U}^\dagger$. 
The two 
possibilities with a single derivative are
\begin{eqnarray}
  \mathbb{V}_{\mu} & = & \frac{i}{2} \left[ \sigma^{\dagger} \partial_\mu
   \sigma + \sigma \partial_\mu \sigma^{\dagger}   \right] \ ,
 \\
  \mathbb{A}_{\mu} & = & \frac{i}{2} \left[ \sigma^{\dagger} \partial_\mu
   \sigma - \sigma \partial_\mu \sigma^{\dagger}   \right] \ .
\end{eqnarray}
$\mathbb{V}_{\mu}$ transforms like a vector field under the $SU(4n)_L\times SU(4n)_R$ 
chiral symmetry and, 
when combined with the derivative, 
can form a 
covariant derivative acting on the heavy-light field or its conjugate: 
\begin{eqnarray}\label{eq:Ddef}
	(H \leftvec D_\mu)_a  = H_b \leftvec D^{ba}_\mu  
	&\equiv& \partial_\mu H_a + i H_b\mathbb{V}_{\mu}^{ba}\ , 
	\nonumber \\
	(\rightvec D_\mu \overline{H})_a  = 
	\rightvec D^{ab}_\mu \overline{H}_b 
	&\equiv& \partial_\mu \overline{H}_a - 
	i \mathbb{V}_{\mu}^{ab} \overline {H}_b\ ,
\end{eqnarray}
with implicit sums over repeated indices.
The covariant derivatives and $\mathbb{A}_\mu$
transform under the chiral symmetry as
\begin{eqnarray}\label{eq:Dtransf}
	H \leftvec D_\mu &\to&  (H \leftvec D_\mu )\mathbb{U}^\dagger\ , \nonumber \\
	\rightvec D_\mu \overline{H} &\to&  \mathbb{U} (\rightvec D_\mu \overline{H})\ ,\nonumber \\
	 \mathbb{A}_\mu &\to&  \mathbb{U} \mathbb{A}_\mu \mathbb{U}^\dagger\ .
\end{eqnarray}

The combined symmetry group of the theory includes Euclidean rotations (or Lorentz symmetry), 
translations, heavy-quark spin, flavor-taste chiral symmetries,
and the discrete
symmetries $C$, $P$, and $T$.   Many of these symmetries are violated by lattice artifacts
and/or light quark masses. Violations to a given order are encoded as spurions in the Symanzik action.
  From these spurions, the heavy-light and light-light fields,
derivatives, the heavy quark 4-velocity $v_\mu$, and the light quark gamma matrix $\gamma_\mu$,
we can construct the chiral Lagrangian and relevant currents order by order.  

Reference~\cite{HL_SCHPT} finds the lowest order heavy-chiral Lagrangian and left-handed current,
as well as higher order corrections. We need primarily the lowest order results here.  For convenience,
we write the Lagrangian in Minkowski space, so that we can make contact with the continuum literature.

We write the leading order (LO) chiral Lagrangian as 
\begin{equation}\label{eq:Lcont}
  \cL_{LO} = \cL_{\rm pion}+ \cL_{\rm HL} 
\end{equation}
where $\cL_{\rm pion}$ is the standard 
\schpt\ Lagrangian \cite{SCHPT} for the light-light mesons, 
and $\cL_{\rm HL}$
is the contribution of the heavy-lights. 
We have\footnote{There is a missing minus sign in Eq.~(35) of Ref.~\cite{HL_SCHPT}.}
\begin{eqnarray}
	\cL_{\rm pion} & = & \frac{f^2}{8} \Tr(\partial_{\mu}\Sigma 
  \partial^{\mu}\Sigma^{\dagger}) + 
  \frac{1}{4}\mu f^2 \Tr(\cM\Sigma+\cM\Sigma^{\dagger})
  \nonumber\\&&{}
  - \frac{2m_0^2}{3}(U_I + D_I + S_I+\ldots)^2 - a^2 \cV \ ,
  \label{eq:Lpion}\\
  -\cV & = & C_1
  \Tr(\xi^{(n)}_5\Sigma\xi^{(n)}_5\Sigma^{\dagger})
  +C_3\frac{1}{2} \sum_{\nu}[ \Tr(\xi^{(n)}_{\nu}\Sigma
    \xi^{(n)}_{\nu}\Sigma) + h.c.] \nonumber \\*&&
  {}+C_4\frac{1}{2} \sum_{\nu}[ \Tr(\xi^{(n)}_{\nu 5}\Sigma
    \xi^{(n)}_{5\nu}\Sigma) + h.c.]
   +C_6\ \sum_{\mu<\nu} \Tr(\xi^{(n)}_{\mu\nu}\Sigma
  \xi^{(n)}_{\nu\mu}\Sigma^{\dagger})  \nonumber \\*&&
   {}+C_{2V}\frac{1}{4} \sum_{\nu}[ \Tr(\xi^{(n)}_{\nu}\Sigma)
   \Tr(\xi^{(n)}_{\nu}\Sigma)
    + h.c.]
   +C_{2A}\frac{1}{4} \sum_{\nu}[ \Tr(\xi^{(n)}_{\nu5}\Sigma)
   \Tr(\xi^{(n)}_{5\nu}\Sigma)
    + h.c.] \nonumber \\*
  && {}+C_{5V}\frac{1}{2} \sum_{\nu} \Tr(\xi^{(n)}_{\nu}\Sigma)
  \Tr(\xi^{(n)}_{\nu}\Sigma^{\dagger})
  + C_{5A}\frac{1}{2}\sum_{\nu}\Tr(\xi^{(n)}_{\nu 5}\Sigma)
  \Tr(\xi^{(n)}_{5\nu}\Sigma^{\dagger}) \label{eq:VSigma} \ ,\\
  \cL_{\rm HL}  &=&  -i \Tr(\overline{H} H v\negcdot \leftvec D )
  + g_\pi \Tr(\overline{H}H\gamma^{\mu}\gamma_5 
  \mathbb{A}_{\mu}) \ .
\label{eq:L-HL}
\end{eqnarray}
Here $\Tr$ denotes a trace over flavor-taste indices and, 
where relevant, 
Dirac indices.  The product  $\overline{H}H$ is treated as a matrix in flavor-taste space: 
$(\overline{H}H)_{ab} \equiv \overline{H}_aH_b$.
The covariant derivative $\leftvec D$ acts only on the field 
immediately preceding it.
For convenience, we work with diagonal fields ($U$, $D$, \dots) and leave
the anomaly ($m_0^2$) term explicit in \eq{Lpion}. 
We can take 
$m^2_0\to\infty$ and go to the physical basis ($\pi^0$, $\eta$, \dots)
at the end of the calculation
\cite{SHARPE_SHORESH}.

To calculate semileptonic form factors, 
we need the chiral representative of the 
left-handed current which destroys a heavy-light meson
of flavor-taste $b$. 
At LO this takes the form 
\begin{equation}\label{eq:LOcurrent}
	j^{\mu,b}_{\rm LO} = \frac{\kappa}{2}\; 
	\trD\bigl(\gamma^\mu \left(1-\gamma_5\right) H\bigr) 
	\sigma^\dagger \lambda^{(b)}\ ,
\end{equation}
where $\lambda^{(b)}$ is a constant vector that fixes the flavor-taste:
$(\lambda^{(b)})_c = \delta_{bc}$, 
and $\trD$ is a trace on Dirac indices only.  

The power counting is a little complicated in the heavy-light case, since many
scales are available. Let $m_q$ be a generic light quark mass, and let $m_\pi^2\propto m_q$ be the corresponding
``pion'' mass, with  $p$ its 4-momentum.  Further, take $k$ as the heavy-light meson's residual momentum.
Then our power counting assumes $k^2 \sim p^2 \sim m_\pi^2 \sim m_q \sim a^2$, where appropriate powers
of the chiral scale or $\Lambda_{QCD}$ are implicit. The leading heavy-light chiral Lagrangian $\cL_{HL}$
is $\cO(k)$, the leading light-light Lagrangian $\cL_{\rm pion}$ is $\cO(p^2, m_q, a^2)$,
and the leading heavy-light current $j^{\mu,b}_{\rm LO}$ is
$\cO(1)$.  Only these leading terms are relevant to the calculation of non-analytic
``chiral logarithms'' at first non-trivial order, which give $\cO(m_q,a^2)$ corrections to leading
expressions for  semileptonic form factors.  

In principle, finding the corresponding analytic corrections requires complete knowledge of
the next-order terms in the Lagrangian and current.
However, since the form factors depend only on the 
the valence and sea quark masses, $a^2$, and the
pion energy in the rest frame of the
$B$ (namely $\vp$),
the form of these corrections is rather
simple and is easily determined 
by the symmetries. The large number of  chiral parameters that can appear 
in higher-order terms in the Lagrangian and
the current collapse down into relatively few free parameters in the form factors.
Unless one wants to write these free parameters in terms of the chiral parameters, complete 
knowledge of the higher-order terms in the Lagrangian and current
is often unnecessary.   
However, one does need to know enough about the higher-order terms to check for the possibility
of relations among the free parameters that multiply different
quantities or that appear in different form factors.   At the order we work here, there is
one relation among the various parameters that determine the linear 
dependence of the two form factors on the valence masses. In order
to be sure that this relation is valid, we need to know all 
terms at next order that can contribute such linear dependence.

Fortunately, all such terms are known. For 
the light-light Lagrangian, \eq{Lpion}, the relevant terms are the standard
$\cO(p^4\!\sim\! m_q^2)$ terms in the continuum \cite{GASSER}. All terms of $\cO(m_qa^2, a^4)$,
which are special to \schpt, are also available \cite{Sharpe:2004is}.
For the heavy-light Lagrangian and current, 
Ref.~\cite{HL_SCHPT} lists all terms which are 
higher order than \eqs{L-HL}{LOcurrent} by a factor of $m_q$ (most important here) or $a^2$.
Reference~\cite{HL_SCHPT} does not attempt a complete catalog of the
terms which are higher than \eqs{L-HL}{LOcurrent} by one or two powers of $k$, 
\ie having one or two derivative insertions.
However, a sufficient number of representative terms of this type are listed to see that
the corresponding free parameters in the form factors are all independent.
We discuss the determination of the analytic terms further in \secref{analytic}.

\section{Generalizing Continuum PQ\chpt\ to \schpt}
\label{sec:PQCHPTtoSCHPT}

We wish to compute the decay $B_x\to P_{xy}$ in \schpt, where $x$ and $y$
are (light) flavor labels.   The taste of
the light quarks in $B$, $P$ and the current also needs to be specified.
We take the $P_{xy}$ to be a ``Goldstone pion"
with taste $\xi_5$.  Let the light quark in the $B$ have taste $\alpha$ ($\alpha=1,\dots,4$);
in flavor-taste notation the light quark has index $a\leftrightarrow x\alpha$.  The current, \eq{LOcurrent},
has flavor-taste $b\leftrightarrow y\beta$. Despite the existence of
taste violations at non-zero lattice spacing, the amplitude turns out to be proportional to
 $(\xi_5/2)_{\alpha\beta}$, with a proportionality factor that is independent of the tastes $\alpha,\beta$.
We will often keep this rather trivial taste-dependence implicit.

In Ref.~\cite{BECIREVIC}, \Bec, \et\ have
calculated the form factors for $B \to \pi$ and $B \to K$ transitions in
continuum PQ\chpt. They assume degenerate sea-quark masses, but leave 
$N_{\rm sea}$, the number of sea quarks, arbitrary. As we explain
below, the $N_{\rm sea}$ dependence is a marker for the underlying
quark flow  \cite{QUARK-FLOW} within the meson diagrams.  Once we have
separated the meson diagrams into their contributions from various
the quark flow diagrams, we can easily generalize the continuum PQ\chpt\ results
to the staggered case, without actually having to calculate any \schpt\ diagrams.
To check our method, however, we have also computed many of the diagrams
directly in \schpt; the results agree.

The key feature that makes possible the generalization of continuum PQ\chpt\ results to \schpt\ results
is the taste-invariance of the leading-order Lagrangian for the heavy-light mesons \cite{HL_SCHPT}.
This means that the continuum vertices and propagators involving
heavy-light mesons 
are trivially generalized to the staggered case: flavor indices (which can
take $N_{\rm sea}$ values if they describe sea quarks) 
simply become flavor-taste indices (taking $4N_{\rm sea}$ sea-quark values).
In one-loop diagrams, taste violations arise only from the light meson (``pion'') 
propagators.  
Propagators and vertices for the staggered and continuum cases are listed Appendix~\ref{app:rules}.

Looking at the expressions in Appendix B of Ref.~\cite{BECIREVIC}, 
we see that there are two types of
terms that can contribute to each
diagram for $B_x\to P_{xy}$: 
a term proportional 
to $N_{\rm sea}$, 
and a term proportional to $1/N_{\rm sea}$.
This is the same behavior that appears, for example, in
light-light \cite{Sharpe:1997by} or heavy-light \cite{Sharpe:1995qp}
 PQ\chpt\ decay constants.

\begin{figure}[htbp]
\begin{center}
\includegraphics[width=6in]{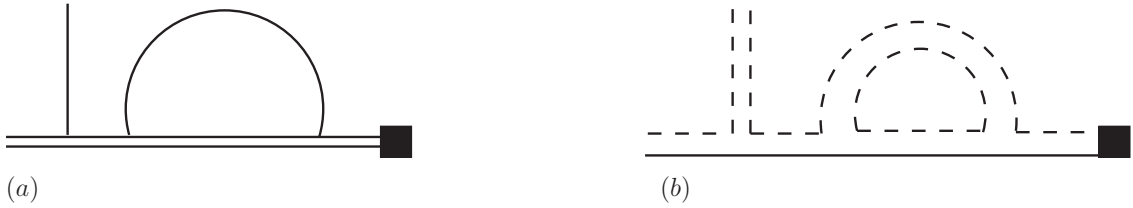}
\caption{Example of a connected one-loop form factor diagram at (a) the meson level and (b) the quark level. 
For the meson diagram, 
the double line is a heavy-light meson while the single line is a pion. 
For the quark-level diagram, 
the solid line is a heavy quark and the dashed line is a light quark. 
The internal sea quark loop is required by the (quark-flow)
connected pion propagator; purely valence
diagrams are only possible with a disconnected pion propagator. Therefore this diagram
gives rise to a factor of $N_{\rm sea}$ in the degenerate case.}
\label{fig:conn_q_lev}
\end{center}
\end{figure}

\begin{figure}[htbp]
\begin{center}
\includegraphics[width=6in]{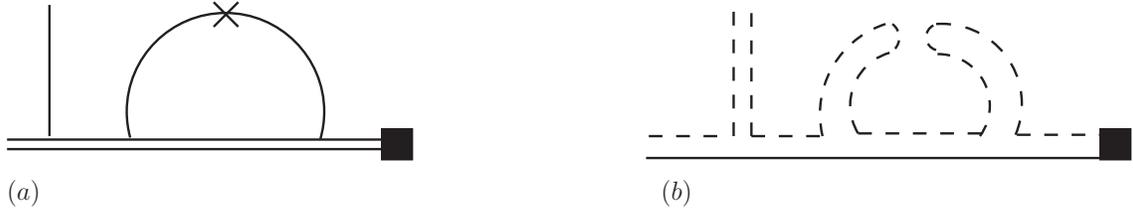}
\caption{Example of a disconnected one-loop form factor diagram at (a) the meson level and (b) the quark level. 
The cross in the meson diagram represents the two-point interactions in \chpt, 
and is represented by the ``hairpin'' in the quark-level diagram. 
There are no factors of $N_{\rm sea}$ but instead factors of $1/N{\rm sea}$ coming from the decoupling of the $\eta'$.}
\label{fig:disc_q_lev}
\end{center}
\end{figure}

The term which is proportional to $N_{\rm sea}$ comes solely from
connected quark-level diagrams, 
an example of which is 
shown in \figref{conn_q_lev} (where (a) is the meson-level diagram and (b) is 
the quark-level diagram).\footnote{By definition, the pion propagator 
in \figref{conn_q_lev}(a) is connected; the version with a disconnected propagator is shown
in \figref{disc_q_lev}(a).}
The appearance
of the quark loop accounts for the factor of $N_{\rm sea}$.  In detail, using 
\eqs{Bstarprop-cont}{B-Bstar-pi-cont}, the loop integrand is proportional to
the connected contraction $\sum_j\bigl\{\Phi_{ij}\Phi_{ji'}\bigr\}_{\rm conn}$, where the
index $j$  is repeated because the heavy-light propagator conserves flavor.
\Eq{PropConnContinuum} then implies that the sum over $j$ produces a factor of $N_{\rm sea}$
when the sea quarks are degenerate.
In the non-degenerate case, there is no factor of $N_{\rm sea}$ but
simply a  sum over the sea-quark flavor of the virtual valence-sea pion.

In the staggered case, the internal heavy propagators, \eqs{Bprop}{Bstarprop}, 
as well as the vertices coupling heavy-light mesons to pions (\eg \eq{B-Bstar-pi}),
 preserve both flavor and taste.  Therefore
\figref{conn_q_lev} is now simply proportional to $\sum_b\bigl\{\Phi_{ab}\Phi_{ba'}\bigr\}_{\rm conn}=
\sum_{j \beta }\bigl\{\Phi_{i\alpha,j\beta}\Phi_{j\beta, i'\alpha'}\bigr\}_{\rm conn}$, where
we have replaced the flavor-taste indices ($a,b,\dots$) with separate flavor ($i,j,\dots$) and
taste ($\alpha,\beta,\dots$) indices.
  From \eq{PropConn}, the loop integrand is then proportional to
\begin{equation}\eqn{taste-average}
\sum_{j,\Xi} 
 \frac{i\delta_{ii'}\delta_{\alpha\alpha'} }
{p^2 - m_{ij,\Xi}^2 + i\epsilon} \ ,
\end{equation} 
where the $\delta_{\alpha\alpha'}$ factor shows that, despite the existence of
taste violations,  the loop preserves the
taste of the light quark in the heavy-light meson and is independent of that taste.

The overall factor of the \schpt\ diagram must be such as to reproduce the continuum result
in the $a\to 0$ limit.  Since pions come in 16 tastes, the sum over pion tastes $\Xi$ in \eq{taste-average}
must come with a factor of $1/16$ compared to the continuum expression.  To see this explicitly,
note first that there are two factors of $1/2$ relative to the continuum coming from the vertices
(compare \eqs{B-Bstar-pi}{B-Bstar-pi-cont}),
due to the non-standard normalization of the taste generators, \eq{T_Xi}.  An additional factor of $1/4$
comes from the \schpt\ procedure for taking into account the fourth root of the staggered determinant: This
is a diagram with a single sea quark loop.

Finally, we need to consider how such a diagram depends on the tastes $\alpha$  and $\beta$ of the
heavy-light meson and the current.
Since the taste indices flow trivially through the heavy-light lines and vertices, and, as we have seen,
through the loops, the taste dependence is simply $(\xi_5/2)_{\alpha\beta}$, where the $\xi_5$ comes from the
outgoing light meson. The factor of $1/2$ is due to the normalization
of the taste generators.

The net result is that terms with factors of $N_{\rm sea}$ in the continuum calculation
of Ref.~\cite{BECIREVIC} are converted to \schpt\ by the rule:
\begin{equation}\label{eq:conn_replace}
  N_{\rm sea} \cF(m^2_M) \to  \frac{(\xi_5)_{\alpha\beta}}{2}\; \frac{1}{16}
  \sum_{f,\Xi} \cF(m^2_{fz,\Xi})
\end{equation}
where
the sum over $f$ is over the sea quark flavors, 
$z$ is the valence flavor flowing through the loop (either $x$ or $y$), $m_M$ 
is the common mass of the $N_{\rm sea}$ mesons
made up of a $z$ valence quark and the degenerate sea 
quarks, 
and $\cF$ is some function
of the pion masses. (For 
heavy-light quantities, 
$\cF$ is often also a function
of the pion energy in the heavy-light rest frame.) 
The masses of pions of various tastes and flavors ($m_{fz,\Xi}$) are given in \eq{pi-masses-specific}.

The terms that are proportional to $1/N_{\rm sea}$ are more subtle.
They arise from diagrams with disconnected pion propagators. The simplest example
is shown at the meson level in
\figref{disc_q_lev}(a) and at the quark level in \figref{disc_q_lev}(b).
The continuum form of the disconnected propagator is given in \eq{DiscContinuum}.
Using the continuum values
$\delta'=m_0^2/3$ and $m_\eta'^2\approx N_{\rm sea}m_0^2/3$, we see that the 
disconnected propagator produces
an overall factor of $1/N_{\rm sea}$ as $m_0\to\infty$.
\Eq{DiscContinuum} can then be written as a 
sum of residues times poles, where the residues
can be rather complicated when the sea masses are non-degenerate (see Appendix~\ref{app:int}). 
Thus, 
the final answer after integration amounts to something of the form
\begin{equation}\label{eq:disc_replace}
	\frac{1}{N_{\rm sea}}\sum_j \hat R_j \tilde \cF(m^2_j)\ ,
\end{equation}
where $\tilde \cF$ is again a general function resulting from the loop integral, 
$\hat R_j$ is the residue of the pole at $q^2=m^2_j$, and 
$j$ ranges over the flavor-neutral mesons involved: the sea mesons, $\pi_0,\eta,\dots$,
and the ``external'' mesons in the disconnected propagator, called
$ii$ and $i'i'$ in \eq{DiscContinuum}. 
When $m_{i'i'}= m_{ii}$, there is a double pole, and \eq{disc_replace} should be replaced
by
\begin{equation}\label{eq:disc_replace_double}
	\frac{1}{N_{\rm sea}}\sum_j \frac{\partial}
	{\partial m_{ii}^2} \left[\hat R_j \tilde \cF(m^2_j)\right]\ ,
\end{equation}
where the sum over $j$ now does not include  $i'i'$.

When the sea quarks are degenerate, the residues simplify considerably. However, by comparing
the general forms in \eqs{disc_replace}{disc_replace_double} to the rather simple terms in
Ref.~\cite{BECIREVIC},
it is easy to move \textit{backwards} from the degenerate case and
determine the form of the expressions  for non-degenerate sea quarks.

The flavor structure in the staggered case is identical to that in the continuum: Flavor remains
a good quantum number, so meson propagators in both cases can only be disconnected
if they are flavor neutral. 
Because of taste violations, however, disconnected
hairpin diagrams can contribute to mesons propagators with three different tastes 
(singlet, vector, and axial vector) at this order in \schpt. 
These three hairpin contributions are quite similar to each other, but there
are a few important differences:
\begin{itemize}
\item[$\bullet$]{} The strength of the hairpin, $\delta'_\Xi$,
depends on the taste $\Xi$ --- see \eq{dp_def}.
\item[$\bullet$]{} In the taste-singlet case, as in the continuum, the hairpin ($4m_0^2/3$) comes
from the anomaly and makes the flavor-singlet meson heavy.
Decoupling the $\eta_I'$ by taking $m_0^2\to\infty$  is therefore a good approximation,
and we do it throughout,
giving rise to an overall factor of $1/N_{\rm sea}$. But in the taste-vector and taste-axial-vector
cases, the hairpins are not particularly large; indeed they are taste-violating effects
that vanish like $a^2$ (up to logarithms) as $a\to0$. So we cannot decouple 
the corresponding mesons, $\eta'_{V}$ and $\eta'_{A}$,
in the taste-vector and taste-axial-vector channels.

\item[$\bullet$]{} The taste matrices associated with the vector and axial-vector
mesons, $\xi_\mu$ and $\xi_{\mu5}$, 
anticommute with the $\xi_5$ coming from the outgoing Goldstone pion.  
Therefore the vector and axial hairpin contributions  will have
an opposite sign from the singlet (and continuum) contribution if the $\xi_5$ needs to be
pushed past a $\xi_\mu$ or $\xi_{\mu5}$ to contract with  the external pion state.
\end{itemize}

\Figref{tree} shows the tree-level diagrams that contribute to the form factors, while \figsref{1loopV}{1loopP} 
show all the non-vanishing one-loop diagrams.
As a first example of the treatment of diagrams with disconnected meson propagators,
consider  \figref{1loopP}(b). It is not hard to see that
this diagram has only a disconnected contribution, shown as a  quark-flow 
diagram in \figref{qu_lev_ex}. 
A connected contribution would require the contraction of the external
light quark fields $x$ and $y$, which make up the outgoing pion.  
That is impossible
since we have chosen $x\not=y$.\footnote{A similar argument will be given
in more detail below when discussing \figsref{qu_lev_ex2disc}{qu_lev_ex2conn}.} 

In our notation, the result
from Ref.~\cite{BECIREVIC} for this diagram in
the continuum partially quenched case with $N_{\rm sea}$ degenerate sea quarks
is: 
\begin{equation}\label{eq:Bec_ex}
	-\frac{g_\pi^2}{(4\pi f)^2}\frac{1}{N_{\rm sea}}
	\left[
	\frac{m^2_Y - m^2_U}{m^2_Y - m^2_X}J^{\rm sub}_1(m_Y,\vp)
	-
	\frac{m^2_X - m^2_U}{m^2_Y - m^2_X}J^{\rm sub}_1(m_X,\vp)
	\right]\ ,
\end{equation}
where $m_U$ is the mass of any of the mesons made up of a sea quark and a sea anti-quark,
$m_X$ and $m_Y$ are the masses of the
flavor-neutral mesons made up of $x\bar x$ and $y\bar y$ quarks, respectively, 
and the function $J_1^{\rm sub}$,  defined below in \eq{sub_J1}, is the result 
of the momentum integral.

The ratios of mass differences in \eq{Bec_ex} can be recognized as the residue
functions  (see Appendix \ref{app:int}) for the various poles. For example,
$(m^2_Y - m^2_U)/(m^2_Y - m^2_X)$ is the residue for the pole at $q^2=m^2_Y$.
These residues are rather simple in this case because of the degeneracy of the
sea quarks. To generalize \eq{Bec_ex} to the completely non-degenerate case,
we simply need to replace the residues by their general expressions.
For $N_{\rm sea}$ non-degenerate sea quarks, 
\eq{Bec_ex} is replaced by 
\begin{equation}\label{eq:Bec_ex2}
	-\frac{g_\pi^2}{(4\pi f)^2}\frac{1}{N_{\rm sea}}
	\sum_j\left[
	\hat R^{[N_{\rm sea}+1, N_{\rm sea}]}_j J^{\rm sub}_1(m_j,\vp)
	\right]\ ,
\end{equation}
where the Minkowski residues $\hat R_j^{[n,k]}$ are defined in \eq{Mink-residues}, and
the sum over $j$ is over the $N_{\rm sea}+1$ mesons that make up the denominator masses 
in the disconnected 
propagator after $m^2_0\to \infty$. (See \eq{DiscContinuum} and the discussion following it.)
We leave implicit, for now, the arguments to the residues in \eq{Bec_ex2}; we will
be more explicit in the final results below.  In addition, we will ultimately express
everything in terms of Euclidean-space residues  $R_j^{[n,k]}$, \eq{residues}, simply because those 
are what have been defined and used previously \cite{SCHPT,HL_SCHPT}.

Cases with double poles present no additional problems, since Ref.~\cite{BECIREVIC}
shows these explicitly as derivatives with respect to squared masses of the
results of single-pole integrals. We will therefore simply get derivatives of the
usual residues, as in \eq{lagrange2}.

As discussed above,
we will need the expression  {\it before}\/ the  $m^2_0\to \infty$ limit is taken in
order to generalize the result to the disconnected taste-vector and axial-vector
cases. 
\Eq{DiscContinuum} and the fact that $m^2_{\eta'}\approx N_{\rm sea}m_0^2/3$ for large $m_0$ allow
us to rewrite \eq{Bec_ex2} as
\begin{equation}\label{eq:Bec_ex3}
	+\frac{g_\pi^2}{(4\pi f)^2}
	\frac{m_0^2}{3}
	\sum_j\left[
	\hat R^{[N_{\rm sea}+2, N_{\rm sea}]}_j J^{\rm sub}_1(m_j,\vp)
	\right]\ .
\end{equation}
The sum over $j$ now includes the $\eta'$. The
sign difference between \eqs{Bec_ex2}{Bec_ex3}
comes from the sign of the mass term in the Minkowski-space $\eta'$ propagator. 

Generalizing \eq{Bec_ex3} to the staggered case is then straightforward.  For the
taste-singlet hairpin contributions, we simply replace each continuum pion mass by the mass
of the corresponding taste-singlet pion.  In other words, we just let $m_j\to m_{j,I}$ in
\eq{Bec_ex3}.  Note that, after the staggered fourth root is properly taken
into account, the taste-singlet $\eta'$ mass goes like $ N_{\rm sea}m_0^2/3$ for large $m_0$, 
as it does in the
continuum, so one could reverse the process that led to \eq{Bec_ex3} and use instead
\eq{Bec_ex2} or even \eq{Bec_ex} (for degenerate sea-quarks),
with $m_j\to m_{j,I}$ in both cases.  Just as for 
diagrams with connected pion propagators (see \eq{conn_replace}),
there is also a trivial overall factor
of $(\xi_5)_{\alpha\beta}/2$, where $\alpha$ and $\beta$ are the tastes of the 
heavy-light meson and the current, respectively, and the $\xi_5$ is due
to the pseudoscalar (Goldstone) taste of the outgoing pion.

For the taste-vector and axial-vector disconnected contributions, a little more
work is required.  We first note that the factor of $m_0^2/3$ in \eq{Bec_ex3} is
simply $\delta'_\Xi/4$ with $\Xi=I$, the strength of the taste-singlet
hairpin,  \eq{dp_def}.\footnote{The factor of $1/4$ just comes
from the different conventional normalization of the generators in the
continuum and staggered cases; 
see Appendix~\ref{app:rules} for further discussion of normalization issues.}
For the other tastes we then replace $\delta'_\Xi$ by the appropriate
hairpin strength  from \eq{dp_def} and also replace the pion masses:  $m_j\to m_{j,\Xi}$.
In addition, there is an overall sign change for this diagram in going from
the singlet to the vector or axial-vector tastes.  This comes from the
fact that the outgoing pion line in \figref{1loopP}(b)
lies between the two ends of the disconnected
propagator.
Using \eq{PropDisc} and the Feynman rules for the heavy-light propagators and vertices
in Appendix~\ref{app:rules}, one sees that   the diagram with a taste-$\Xi$ disconnected
propagator goes like $ \big(\, T_\Xi \,\xi_5 \,T_\Xi\,\big)_{\alpha\beta}$.
This leads to a positive sign for $\Xi=I$ but a negative sign for tastes that
anticommute with $\xi_5$.  Finally, the fact that there are 
four degenerate taste-vector (or axial-vector) pions at this order
leads to an additional overall factor of four. 

When we attempt to apply the same procedure to the other diagrams in  \figsref{1loopV}{1loopP},
we find a further complication in diagrams \figref{1loopV}(a), \figref{1loopV}(b),
and \figref{1loopP}(c), where the external pion and one or more internal pions emerge
from the same vertex.  The problem is that the ordering
of the taste matrices at the vertex is not determined by the meson-level diagram
(\ie each diagram can correspond to several orderings), so we do not immediately
know the relative sign of taste-vector and axial-vector  contributions
relative to the singlet contribution.  Nevertheless, a quark-flow analysis allows us
to identify appropriate ``flags'' that signal which terms in Ref.~\cite{BECIREVIC}
come from which orderings at the vertex.

As an example of the procedure in this case, consider \figref{1loopP}(c).  The corresponding
quark flow diagrams with disconnected pion propagators are shown \figref{qu_lev_ex2disc}.
In \figref{qu_lev_ex2disc}(a), the outgoing pion lies between the two ends
of the disconnected propagator. This produces a change in sign of the taste-vector
and axial-vector hairpin contributions relative to the taste-singlet one, just as for
\figref{1loopP}(b).  In \figref{qu_lev_ex2disc}(b), on the other hand, the outgoing
pion is emitted outside the disconnected propagator, and all the hairpin
contributions have the same sign.  The same is true of the reflected version of
\figref{qu_lev_ex2disc}(b), which has the outgoing pion emerging
from the other side of the vertex.  

Fortunately,  Figs.~\ref{fig:qu_lev_ex2disc}(a) 
and (b) are distinguished 
by their flavor structure, even in the continuum.  In \figref{qu_lev_ex2disc}(a),
the two ``external'' mesons in the disconnected propagator have different flavors:
The one on the left is an $X$ meson (an $x\bar x$ bound state); while the one
on the right is a $Y$ meson (a $y\bar y$ bound state).  In  \figref{qu_lev_ex2disc}(b),
both external mesons in the disconnected propagator are $Y$ mesons. Similarly, the
reflected version of \figref{qu_lev_ex2disc}(b) has two $X$ mesons in the disconnected propagator.
This flavor structure is immediately apparent in the results of Ref.~\cite{BECIREVIC}.
The parts of  \figref{1loopP}(c) that come from the quark flow of \figref{qu_lev_ex2disc}(a)
are proportional to the function called $H_1$, which depends on the masses $m_X$ and $m_Y$
(in our notation), as well as the sea-meson mass. 
The parts of  \figref{1loopP}(c) that come from the quark flow of \figref{qu_lev_ex2disc}(b)
(or its reflected version) are proportional to the function called $G_1$, which depends only 
on the mass $m_Y$ (or $m_X$) and the sea-meson mass. To generalize the results of 
Ref.~\cite{BECIREVIC} to the staggered case, we thus can use the method outlined
above, and simply include an extra minus sign for those taste-vector and axial-vector hairpin
contributions proportional to $H_1$ (relative to the taste-singlet contributions), but not
for those proportional to $G_1$.  This approach also works for the other problematic
diagrams,  Figs.~\ref{fig:1loopV}(a) and (b).  

The reader may wonder why the complication associated with ordering the taste matrices
at the vertices does not occur when the internal pion propagator is connected,
but only in the disconnected case.    \Figref{qu_lev_ex2conn} shows possible
quark-flow diagrams for \figref{1loopP}(c) with a connected pion propagator.
\Figref{qu_lev_ex2conn}(a) cannot occur in our case because we have assumed
that $x$, the light flavor of the heavy-light meson, is different from $y$,
the light flavor of the weak current.\footnote{Equivalently, we have assumed that
the outgoing pion is flavor charged.} The same reasoning is what allows us to
rule out any connected contributions to \figref{1loopP}(b), as mentioned above.
Thus all contributions with connected propagators
are of the type shown in  \figref{qu_lev_ex2conn}(b), or its reflected version, and these
never have a sign difference between terms with different internal pion tastes.

We note that one can
reproduce the  \schpt\ results for 
light-light \cite{SCHPT} and heavy-light \cite{HL_SCHPT} mesons
by starting from the continuum PQ\chpt\ in Refs.~\cite{Sharpe:1997by} or \cite{Sharpe:1995qp},
respectively, and following the procedure described above.  The computations
are in fact slightly more difficult in those cases than in the one at hand, because
Refs.~\cite{Sharpe:1997by} and \cite{Sharpe:1995qp} do not explicitly separate
double-pole from single-pole contributions.  
It is therefore takes a little work to express the answers from those references in the
form of our residue functions, which is the necessary first step before generalizing
to the staggered case. 

\section{Form Factors for $B\to P$ Decay}\label{sec:form}

The standard form factor decomposition for the matrix element between a
$B_x$ meson and a $P_{xy}$ meson is
\begin{equation}\eqn{formfac-rel}
	\left\langle P_{xy}(p) | \bar y \gamma_\mu b | B_x(p_B)\right
	\rangle = \left[ (p_B + p)_\mu - q_\mu \frac{m^2_{B_x}
	-m^2_{P_{{xy}}}}{q^2}\right]F_+(q^2) +
	\frac{m^2_{B_x}-m^2_{P_{{xy}}}}{q^2}q_\mu F_0(q^2)\ ,
\end{equation}
where $q = p_B - p$ is the momentum transfer. We are suppressing
taste indices everywhere, but emphasize that 
the light pseudoscalar $P_{xy}$ is assumed to be the Goldstone meson
(taste $\xi_5$).
In the heavy quark limit, it is more convenient
to write this in terms of form factors which are independent of the heavy
meson mass
\begin{equation}\eqn{formfac-HQET}
	\left\langle P_{xy}(p) | \bar y \gamma_\mu b | B_x(v)\right
	\rangle_{HQET}
	= \left[ p_\mu - (\vp)v_\mu \right]f_p(\vp) +
	v_\mu f_v(\vp)\ ,
\end{equation}
where $v$ is the four-velocity of the heavy quark, and \vp\ is the energy of
the pion in the heavy meson rest frame. 
Recall that the
QCD heavy meson state and the HQET heavy meson state are related by
\begin{equation}
	| B(p_B)\rangle_{QCD} = \sqrt{m_B}| B(v)\rangle_{HQET}\ .
\end{equation}
The form factors $f_p$ and
and $f_v$ are often called $f_\perp$ and $f_\parallel$, respectively.
As discussed in \secref{PQCHPTtoSCHPT},
the taste indices are left implicit in 
\eqs{formfac-rel}{formfac-HQET}, as are the trivial overall factors of $(\xi_5/2)_{\alpha\beta}$
in the matrix elements.

\begin{figure}[htbp]
\begin{center}
\includegraphics[width=6in]{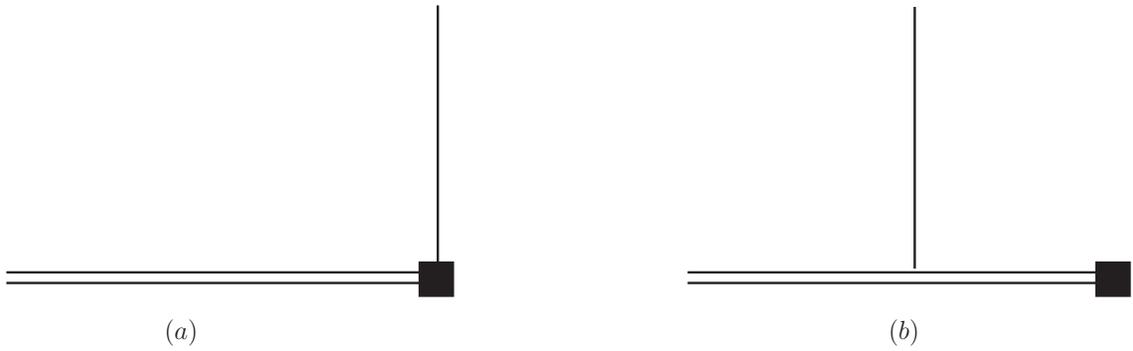}
\caption{Tree level diagrams for (a) $f_v$ and (b) $f_p$. 
The double line is the heavy-light meson and the single line is the pion.}
\label{fig:tree}
\end{center}
\end{figure}

\begin{figure}[htbp]
\begin{center}
\includegraphics[width=6in]{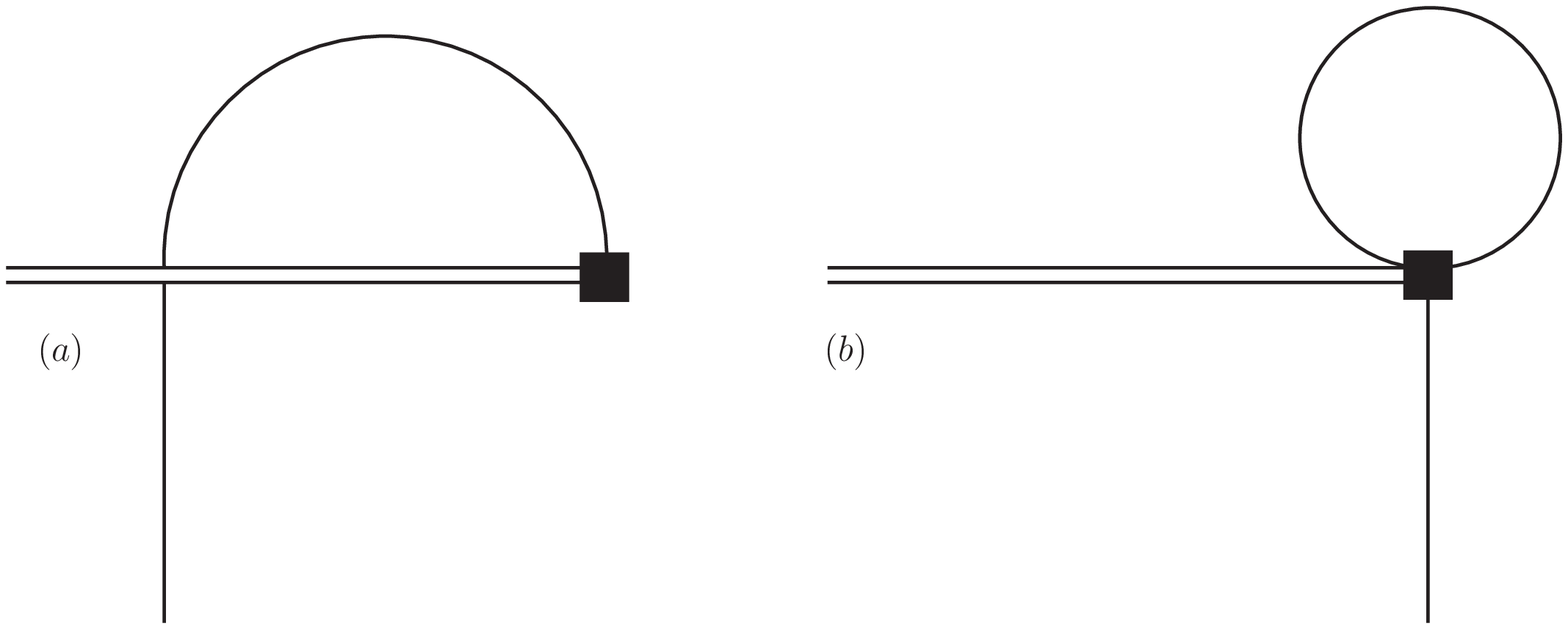}
\caption{One-loop $f_v$ diagrams. The internal light meson lines may
in general be connected or disconnected: possible hairpin insertions
are not shown explicitly.}
\label{fig:1loopV}
\end{center}
\end{figure}

The tree-level diagrams for $B_x\to P_{{xy}}$ are shown in \figref{tree}.
\Figref{tree}(a) is the tree-level ``point'' contribution to $f_v$, while
\figref{tree}(b) is the tree-level ``pole'' contribution to $f_p$. We 
have
\begin{equation}\eqn{fvptree}
	f^{\rm tree}_v(\vp) = \frac{\kappa}{f}\ , \quad
	f^{\rm tree}_p(\vp) = \frac{\kappa}{f}
	\frac{g_\pi}{\vp + \Delstar}\ ,
\end{equation}
where $\Delstar = m_{B^*} - m_{B}$ is the mass difference of the
vector and pseudoscalar heavy-light meson masses at leading order in the
chiral expansion, \ie neglecting all effects of light-quark masses. As
in Refs.~\cite{FALK,BECIREVIC}, we drop this splitting inside loops, but keep
it in the internal $B^*$ line in the tree-level diagram \figref{tree}(b).
This forces the tree-level
pole in $f_p$ to be at $m_{B^*}$, the physical point.  
Dropping $\Delstar$ inside loops is consistent at leading order
in HQET, which is the order to which we are working. It would also be 
consistent, parametrically, to drop $\Delstar$ everywhere. But this would
not be convenient, since the $m_{B^*}$ pole is physically important for $f_p$.

\begin{figure}[htbp]
\begin{center}
\includegraphics[width=6in]{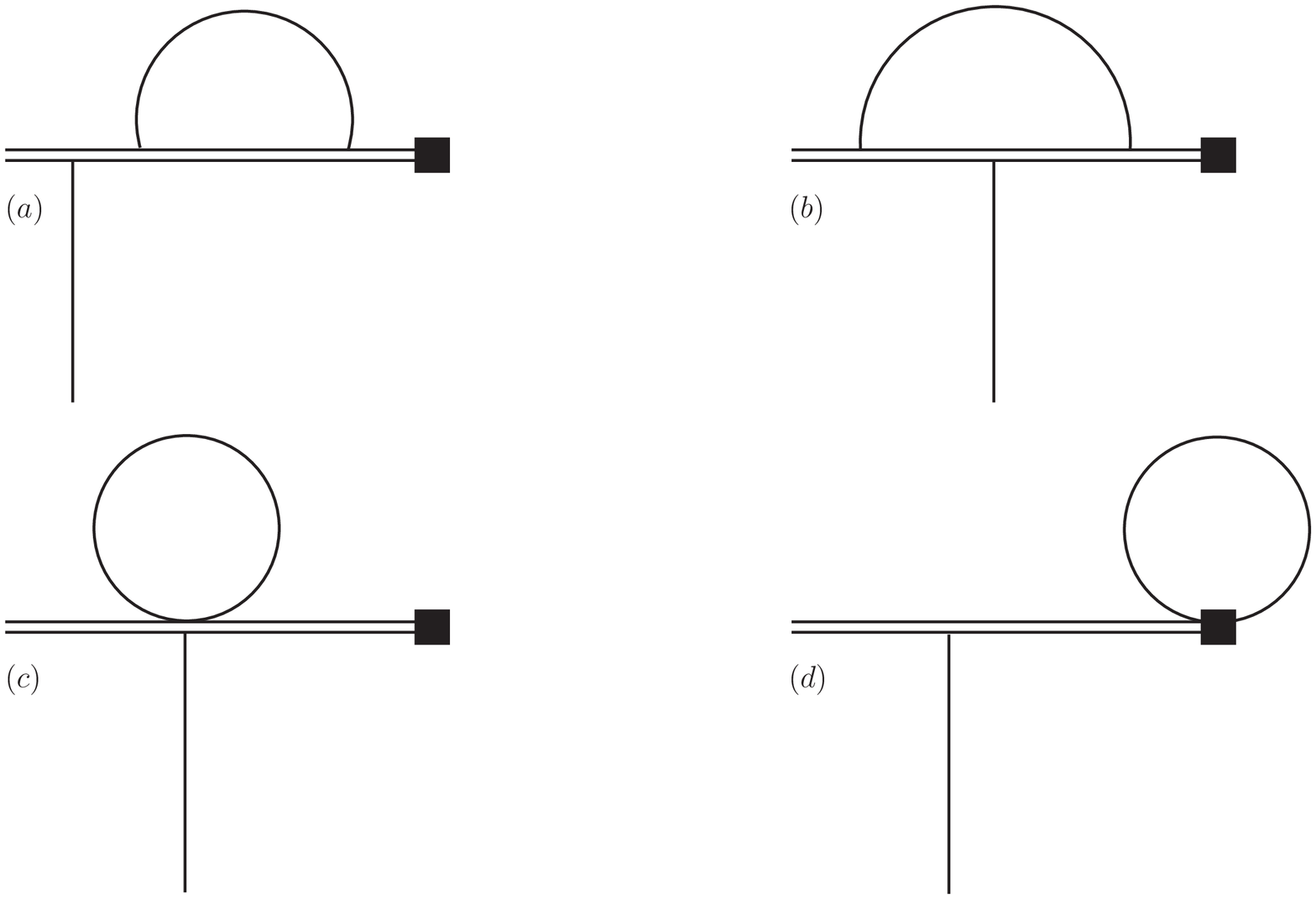}
\caption{One-loop $f_p$ diagrams. The internal light meson lines may
in general be connected or disconnected: possible hairpin insertions
are not shown explicitly.}
\label{fig:1loopP}
\end{center}
\end{figure}

\begin{figure}[htbp]
\begin{center}
\includegraphics[width=4in]{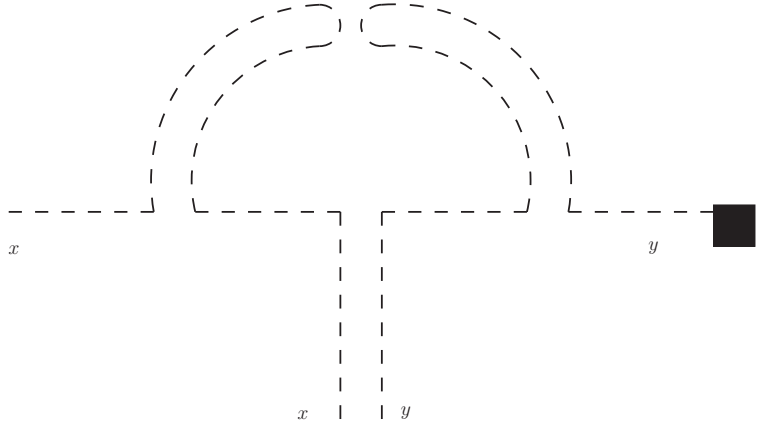}
\caption{The quark-flow diagram for \figref{1loopP}(b),
omitting the heavy quark line for clarity.
The mesons in the loop are $X$ and $Y$ mesons, 
flavor-neutral mesons made up of $x$ and $y$ quarks. 
Note that even though only a single hairpin insertion is shown explicitly, the figure
should be interpreted as representing all diagrams with one or more hairpins.}
\label{fig:qu_lev_ex}
\end{center}
\end{figure}

\begin{figure}[htbp]
\begin{center}
\includegraphics[width=5in]{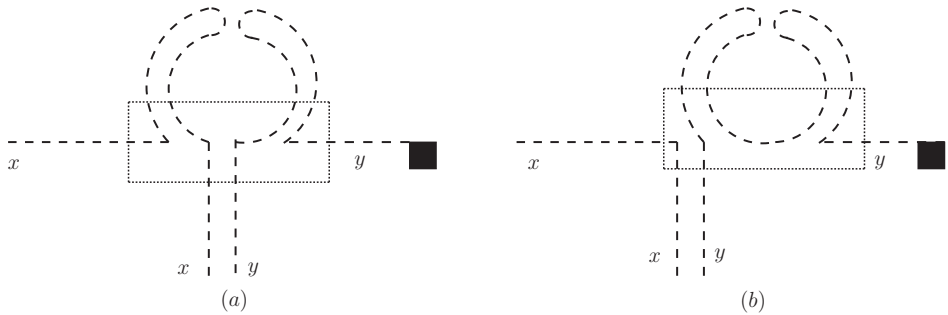}
\caption{Possible quark-flow diagrams for \figref{1loopP}(c) with a disconnected
meson propagator in the loop. The solid rectangle encloses
the 5-point vertex of  \figref{1loopP}(c). The heavy quark line has been omitted for clarity.
A ``reflected'' version of diagram (b), with
the outgoing pion on the other side of the vertex, is also possible.}
\label{fig:qu_lev_ex2disc}
\end{center}
\end{figure}

\begin{figure}[htbp]
\begin{center}
\includegraphics[width=5in]{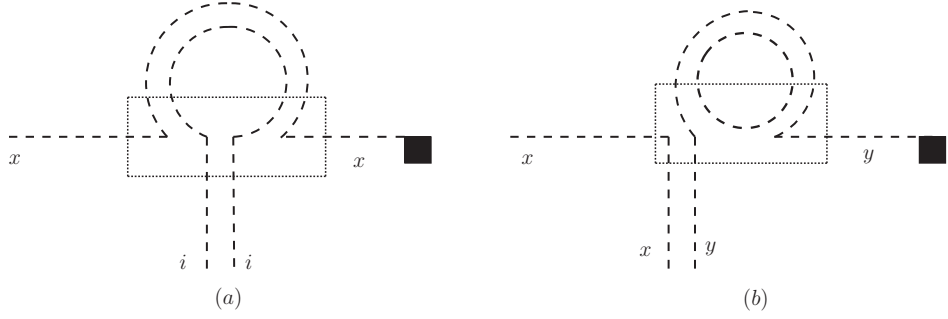}
\caption{Possible quark-flow diagrams for \figref{1loopP}(c) with a connected
meson propagator in the loop. The solid rectangle encloses
the 5-point vertex of  \figref{1loopP}(c). The heavy quark line has been omitted for clarity.
Since we have assumed that $x$ and $y$ are different flavors, diagram (a) cannot
occur in our case. Diagram (b) can occur,
as can  a ``reflected'' version  with the outgoing pion on the other side of the vertex.}
\label{fig:qu_lev_ex2conn}
\end{center}
\end{figure}

The non-zero diagrams that correct the form factors to one loop are shown in
\figref{1loopV} for $f_v$ and \figref{1loopP} for $f_p$. 
\tabref{bec_us} lists the correspondences
between these diagrams and those of Ref.~\cite{BECIREVIC}.
A number of other
diagrams, which can arise in principle, vanish identically due to the
transverse nature of the $B^*$ propagator, \eq{Bstarprop};
these additional diagrams can be
found in Ref.~\cite{BECIREVIC}. 
We do not indicate hairpin vertices explicitly in \figsref{1loopV}{1loopP};
the internal pion propagators in these diagrams may be either connected  or disconnected.

Before generalizing the results in Ref.~\cite{BECIREVIC} to \schpt, we
discuss a subtle issue  that affects \figref{1loopP}(a).  
If the splitting $\Delstar$ is dropped on internal  $B^*$ lines in loop
diagrams, as is done in Ref.~\cite{BECIREVIC}, this diagram has a
spurious singularity (a double pole) at $v\cdot p=0$, the edge of the 
physical region. 
The singularity arises from the presence of the two  $B^*$ lines that 
are not inside the loop integral and therefore can be on mass shell in the absence of 
$B^*$-$B$ splitting.
Including $\Delstar$ on all such internal ``on-shell''
$B^*$ lines (\ie lines not inside the loops themselves), as is done
in Ref.~\cite{FALK}, at least pushes
the unnatural double pole out of the  physical region. We will follow this
prescription for including the splitting, but take it one step further.  The loop
in \figref{1loopP}(a) is a self-energy correction on the internal $B^*$
line.  The double pole results from not iterating the self-energy 
and summing the geometric series.  We will follow
the more natural course and sum the series; doing so restores a standard single-pole 
singularity.  

There is a further one-loop contribution that can naturally be included in \figref{1loopP}(a).
The corresponding tree-level graph, \figref{tree}(b), gets two kinds of corrections that are not
shown in \figref{1loopP}. One comes simply from the wavefunction renormalizations on
the external pion and $B$ lines; we include those terms explicitly below.  The second contribution
arises from the one-loop shift in the external meson mass.  
Since this mass shift depends on the
flavor of the light quark in the $B_x$, namely $x$, we call it $\delta M_x = \Sigma_x(v\cdot k=0)$,
with $\Sigma_x(v\cdot k)$ the self-energy for $B_x$ or $B^*_x$. Note 
that $\Sigma_x$ is the same for both the $B_x$ and the $B^*_x$, since the splitting
$\Delstar$ is dropped inside loops. When the external $B_x$
line in \figref{tree}(b) is put on mass-shell at one loop, the denominator
of the internal $B_y^*$ propagator changes from 
$-2(\vp + \Delstar)$ to  $-2(\vp + \Delstar-\delta M_x)$.
It is convenient to define $ \Delstar_{yx}$ as the full splitting between a $B^*_y$
and a $B_x$:
\begin{equation}\eqn{Deltaxy}
 \Delstar_{yx} \equiv M_{B^*_y} - M_{B_x} =  \Delstar + \delta M_y - \delta M_x
\end{equation}
The internal $B_y^*$ propagator now becomes
 $-2(\vp + \Delstar_{yx}-\delta M_y)$. The contribution from the mass shift may then be 
combined with the tree-level and \figref{1loopP}(a) contributions to give:
\begin{eqnarray}
	f^{\rm self}_p(\vp) &=& \frac{\kappa}{f}\;\,
	\frac{g_\pi}{\vp + \Delstar_{yx} + D(\vp)}\ ,\eqn{fpself}\\
	D(\vp) &\equiv & \Sigma_y(\vp) - \Sigma_y(0)\ , \eqn{D}
\end{eqnarray}
where the subtraction in $D$ comes from the effect of putting the $B_x$
on mass shell, \via\ \eq{Deltaxy}.

The main difference between the approach taken to the spurious
singularity of \figref{1loopP}(a) and that of  \Bec\ \et\ \cite{BECIREVIC} 
is that they work to first order in the self-energy in the 
corresponding diagram (their  diagram (7)). 
Expanding \eq{fpself}, we find that $D$ is related in the continuum
limit to what Ref.~\cite{BECIREVIC} calls $\delta f_p^{(7)}$ by
\begin{equation}\eqn{Drelation}
	D(\vp)  =  -\vp\;\; \delta f_p^{(7)}\ .
\end{equation}
Thus we can find the staggered $D(\vp)$ simply by applying the methods
of \secref{PQCHPTtoSCHPT} to $\delta f_p^{(7)}$.

We can now write down the expressions for the form factors for
$B_x\to P_{xy}$ decay. For the point form factor, $f_v$, we have 
\begin{eqnarray}\label{eq:fv}
	f_v^{B_x\to P_{xy}}  &=&
	f^{\rm tree}_v\bigl[
	1 + \delta f_v^{B_x\to P_{xy}} + c_x^v m_x + c_y^v m_y +
	c_{\rm sea}^v (m_u + m_d + m_s)
	\nonumber\\&&{}+ c_1^v (\vp) + c_2^v (\vp)^2
	 + c_a^v a^2
	\bigr]\ ,
\end{eqnarray}
where $ f^{\rm tree}_v$ is given by \eq{fvptree}, and
the analytic coefficients $c^v_x,\; c^v_y,\; \dots$ arise from next-to-leading order (NLO) terms in
the heavy-light chiral Lagrangian (see \secref{analytic}). The non-analytic pieces, which come
from the diagrams shown in \figref{1loopV} as well as the 
wavefunction renormalizations, are included in
$\delta f_v^{B_x\to P_{xy}}$:
\begin{equation}\label{eq:deltfv}
	\delta f_v^{B_x\to P_{xy}} = \delta f_v^{\ref{fig:1loopV}(a)} +
	\delta f_v^{\ref{fig:1loopV}(b)}
	+\frac{1}{2}\delta Z_{B_x}+
	\frac{1}{2}\delta Z_{P_{xy}} \ .
\end{equation}
The wavefunction renormalization terms, $\delta Z_{B_x}$ and $\delta Z_{P_{xy}}$, 
have been calculated previously \cite{SCHPT,HL_SCHPT} in
\schpt\ and are listed in Appendix~\ref{app:wf_ren}. 

For the $f_p$ form factor, we write
\begin{eqnarray}\label{eq:fp}
	f_p^{B_x\to P_{xy}} & = &
	f_p^{\rm self} +
	\tilde f^{{\rm tree}}_p\bigl[
	\delta f_p^{B_x\to P_{xy}} + c_x^p m_x + c_y^p m_y +
	c_{\rm sea}^p (m_u + m_d + m_s)
	\nonumber\\* &&{}+ c_1^p (\vp)
	+ c_2^p (\vp)^2 + c_a^p a^2
	\bigr]\ .
\end{eqnarray}
where $f_p^{\rm self}$ is defined in \eq{fpself}, and
\begin{equation}\eqn{fptreetilde}
	\tilde f^{\rm tree}_p(\vp) \equiv \frac{\kappa}{f}\;\,
	\frac{g_\pi}{\vp + \Delstar_{xy}}\ .
\end{equation}

Non-analytic contributions are summarized in the function
$D(\vp)$ in $f_p^{\rm self}$, \eq{D}, and $\delta f_p^{B_x\to P_{xy}}$, which comes from
Figs.~\ref{fig:1loopP}(b)-(d) and wavefunction renormalizations. 
Explicitly,
\begin{equation}\label{eq:deltfp}
	\delta f_p^{B_x\to P_{xy}} =
	\delta f_p^{\ref{fig:1loopP}(b)} +
	\delta f_p^{\ref{fig:1loopP}(c)} +
	\delta f_p^{\ref{fig:1loopP}(d)} +
	\frac{1}{2}\delta Z_{B_x} +
	\frac{1}{2}\delta Z_{P_{xy}} \ .
\end{equation}
For simplicity, we do not
include the superscript $B_x\to P_{xy}$ on the individual diagrams in \eqs{deltfv}{deltfp}.

Using $\tilde f^{{\rm tree}}_p$, which includes the
full $B^*_y$--$B_x$ splitting $\Delstar_{yx}$, rather than
$f^{{\rm tree}}_p$, \eq{fvptree}, changes \eq{fp} only by higher-order terms.
However, it is convenient
to keep the same splitting in both $f_p^{\rm self}$ and the other terms in
\eq{fp}.
Note that it is also consistent at this order to use the alternative form
\begin{eqnarray}\label{eq:fp_other}
	f_p^{B_x\to P_{xy}} & = &
	f_p^{\rm self} \bigl[1 +
	\delta f_p^{B_x\to P_{xy}} + c_x^p m_x + c_y^p m_y +
	c_{\rm sea}^p (m_u + m_d + m_s)
	\nonumber\\* &&{}+ c_1^p (\vp)
	+ c_2^p (\vp)^2 	 + c_a^p a^2
	\bigr]\ ,
\end{eqnarray}

The analytic terms in $f_v$ and $f_p$ are not all independent.  As mentioned in \secref{hqetschpt}, 
there is one relation among the terms that control the valence mass dependence:
\begin{equation}\eqn{relation}
c_x^p + c_x^v  = c_y^p + c_y^v 
\end{equation}
We show that this relation follows from the higher order terms in the Lagrangian and
current in \secref{analytic}.  All other NLO parameters in \eqs{fv}{fp} are independent.

\subsection{Form factors for 3-flavor partially quenched \schpt}\label{sec:PQschpt}

First we display the results for the
individual diagrams shown in \figsref{1loopV}{1loopP} for the fully
non-degenerate
case with three dynamical flavors (the ``\opopo'' case). This means that we
have already taken into account the transition from $4$ to $1$
tastes per flavor. Indeed, our method of generalizing
the partially quenched continuum expressions to the staggered case automatically
includes this adjustment.
We detail below the minor changes needed to obtain
$2\!+\!1$ results from those in the  \opopo\ case. 

We first define sets of masses which appear in the numerators and
denominators of the disconnected propagators with taste labels implicit (see
Appendix~\ref{app:int}):
\begin{eqnarray}
  \mu^{(3)} & = & \{m^2_U,m^2_D,m^2_S\}\label{eq:num_masses}\ ,\\*
  \cM^{(3,x)} & = & \{m_X^2,m_{\pi^0}^2, m_{\eta}^2\}
  \label{eq:den_masses_xI}\ ,\\*
  \cM^{(4,x)} & = & \{m_X^2,m_{\pi^0}^2, m_{\eta}^2, m_{\eta'}^2\}
  \label{eq:den_masses_xA}\ , \\*
  \cM^{(4,xy)} & = & \{m_X^2,m_Y^2,m_{\pi^0}^2, m_{\eta}^2\}
  \label{eq:den_masses_xyI}\ ,\\*
  \cM^{(5,xy)} & = & \{m_X^2,m_Y^2,
  m_{\pi^0}^2, m_{\eta}^2, m_{\eta'}^2\}
  \label{eq:den_masses_xyA}\ .
\end{eqnarray}
For the mass sets \eqref{eq:den_masses_xI} and \eqref{eq:den_masses_xA},
there are also corresponding sets with $x\to y$ and $X\to Y$. When we
show explicit taste subscripts such as $I$ or $V$ on the mass sets
$\mu$ or $\cM$, it means that all the masses in the set have that taste.

The functions that appear in the form factors are\footnote{For ease of comparison to
Ref.~\cite{BECIREVIC}, we use $I_1(m)$ instead of $\ell(m^2)$ (as in
Refs.~\cite{SCHPT,HL_SCHPT}) for the chiral logarithm.}
\begin{eqnarray}
	I_1(m)  & = & m^2 \ln \left(\frac{m^2}{\Lambda ^2}\right)
	\label{eq:I1}\ , \\
	I_2(m,\Delta)  & = & -2\Delta^2
	\ln \left(\frac{m^2}{\Lambda^2}\right)
	-4\Delta^2 F\left(\frac{m}{\Delta}\right)
	+2\Delta^2 \label{eq:I2}\ , \\
	J_1(m,\Delta)  & = &
	\left(-m^2 + \frac{2}{3}\Delta^2\right)
	\ln\left(\frac{m^2}{\Lambda^2}\right)
	+\frac{4}{3}(\Delta^2-m^2)
	F\left(\frac{m}{\Delta}\right)-
	\frac{10}{9}\Delta^2
	+\frac{4}{3}m^2  \label{eq:J1}\ , \\
	F(x) & = & \left\{
	\begin{aligned}
	{\sqrt{1-x^2}}\;\; &
	\tanh^{-1}\left( \sqrt{1-x^2}\right), \,
	&0\le x\le 1\\
	-\sqrt{x^2-1}\;\; &
	\tan^{-1} \left( \sqrt{x^2-1}\right),\, &x\ge 1\; .
	\end{aligned}\label{eq:Fdef}\right.
\end{eqnarray}
The main difference in these formulae with those of \Bec\ \et\ \cite{BECIREVIC} is
that they keep the divergence pieces, while
we have renormalized  as in Refs.~\cite{SCHPT,HL_SCHPT}. 
To convert to our form, replace the $\msbar$ scale $\mu$ in
Ref.~\cite{BECIREVIC} with the chiral scale $\Lambda$ and set their
quantity  $\bar\Delta$ to zero, where
\begin{equation}
	\bar\Delta \equiv \frac{2}{4-d} - \gamma + \ln(4\pi) + 1 \ ,
\end{equation}
with $d$ the number of dimensions.  
$F(x)$ is only
needed for positive $x$; so we use the simpler
form given in 
Ref.~\cite{FALK},  rather than the more general version worked out in Ref.~\cite{STEWART}
and quoted in  Ref.~\cite{BECIREVIC}. We do not list the function $J_2$, which appears in the integral
$ \cJ^{\mu\nu} $ of \eq{J1J2int} but does not enter the final answers.

We also define a ``subtracted'' $J_1$ function by
\begin{equation}\label{eq:sub_J1}
	J_1^{\rm sub}(m,\Delta) \equiv J_1(m,\Delta)
	- \frac{2\pi m^3}{3\Delta}\ .
\end{equation}
The subtraction term cancels the singularity when $\Delta\to 0$.
The function $ J_1^{\rm sub}$ enters naturally in the expression for the
self energy correction $D(\vp)$ because of the the subtraction in \eq{D}.
It also turns out to arise from the integral in \figref{1loopP}(b) --- see
Eq.~(26) in Ref.~\cite{FALK}.

For the point corrections in the \opopo\ case, we have
\begin{eqnarray}
  \left(\delta f_v^{\ref{fig:1loopV}(a)}
  \right)^{B_x\to P_{xy}}_{1\!+\!1\!+\!1}
  & = & \frac{1}{2 (4 \pi f)^2} \Biggl\{
  \frac{1}{16}\sum_{f,\Xi}
  \Bigl[I_1(m_{yf,\Xi}) + 2 I_2(m_{yf,\Xi},\vp)\Bigr]
  \nonumber\\&&{}
  +\frac{1}{3}\Biggl[\sum_{j\in \cM^{(4,xy)}}
  R^{[4,3]}_{j}\left(\cM^{(4,xy)}_I ; \mu^{(3)}_I\right)
    \left[I_1(m_{j,I}) + 2I_2(m_{j,I},\vp)\right]
    \nonumber\\&&+ \frac{\partial}{\partial m_{Y,I}^2} \bigg(
    \sum_{j\in \cM^{(3,y)}}
    R^{[3,3]}_{j}\left(\cM^{(3,y)}_I ; \mu^{(3)}_I\right)
    \left[I_1(m_{j,I}) + 2I_2(m_{j,I},\vp)\right]\bigg)
    \Biggr]
  \nonumber \\ &&{}
  +a^2\delta'_V\Biggl[\frac{\partial}{\partial m_{Y,V}^2} \bigg(
    \sum_{j\in \cM^{(4,y)}}
    R^{[4,3]}_{j}\left(\cM^{(4,y)}_V ; \mu^{(3)}_V\right)
    \left[I_1(m_{j,V}) + 2I_2(m_{j,V},\vp)\right]\bigg)
    \nonumber\\&&-\sum_{j\in \cM^{(5,xy)}}
    R^{[5,3]}_{j}\left(\cM^{(5,xy)}_V ; \mu^{(3)}_V\right)
    \left[I_1(m_{j,V}) + 2I_2(m_{j,V},\vp)\right]
    \Biggr] \nonumber\\&&{}+[V\to A]
 \Biggr\} \ , \eqn{fv4a}
  \\
%%%%%%%%%%%%%%%%%%%%%%%%%%%%%%%%%%%%
 \left( \delta f_v^{\ref{fig:1loopV}(b)}
  \right)^{B_x\to P_{xy}}_{1\!+\!1\!+\!1}
  &=&-\frac{1}{6 (4 \pi f)^2}\Biggl\{
  \frac{1}{16}\sum_{f,\Xi}
  \left[I_1(m_{xf,\Xi}) + I_1(m_{yf,\Xi}) \right]
  \nonumber\\&&{}
  +\frac{1}{3}\biggl[ \frac{\partial}{\partial m_{Y,I}^2}\bigg(
    \sum_{j\in \cM^{(3,y)}}
    R^{[3,3]}_{j}\left(\cM^{(3,y)}_I ; \mu^{(3)}_I\right)
    I_1(m_{j,I})\bigg) \nonumber\\&&{}
    + \frac{\partial}{\partial m_{X,I}^2}\bigg(
    \sum_{j\in \cM^{(3,x)}}
    R^{[3,3]}_{j} \left(\cM^{(3,x)}_I ; \mu^{(3)}_I\right)
    I_1(m_{j,I}) \bigg) \nonumber\\&&{}
    -\sum_{j\in \cM^{(4,xy)}}
    R^{[4,3]}_{j}\left(\cM^{(4,xy)}_I ; \mu^{(3)}_I\right)
    I_1(m_{j,I})
     \biggr]\nonumber\\&&{}
  +a^2\delta'_V\biggl[\frac{\partial}{\partial m_{Y,V}^2} \bigg(
    \sum_{j\in \cM^{(4,y)}}
    R^{[4,3]}_{j}\left(\cM^{(4,y)}_V ; \mu^{(3)}_V\right)
    I_1(m_{j,V}) \bigg) \nonumber\\&&{}
    + \frac{\partial}{\partial m_{X,V}^2} \bigg(
    \sum_{j\in \cM^{(4,x)}}
    R^{[4,3]}_{j}\left(\cM^{(4,x)}_V ; \mu^{(3)}_V\right)
    I_1(m_{j,V}) \bigg) \nonumber\\&&{}
      +\sum_{j\in \cM^{(5,xy)}}
      R^{[5,3]}_{j} \left(\cM^{(5,xy)}_V ; \mu^{(3)}_V\right)
    I_1(m_{j,V})
      \biggr]+[V\to A]
  \Biggr\}  \ . \eqn{fv4b}
\end{eqnarray}

Those that correct the pole form factors are
\begin{eqnarray}
  \left(D \right)^{B_x\to P_{xy}}_{1\!+\!1\!+\!1}
  &=& - \frac{3g_\pi^2\vp}{(4 \pi f)^2}
  \Biggl\{ \frac{1}{16}\sum_{f,\Xi}
  J_1^{\rm sub}(m_{yf,\Xi}, \vp)\nonumber \\ &&{}
  + \frac{1}{3}\sum_{j\in \cM^{(3,y)}}
  \frac{\partial}{\partial m_{Y,I}^2}\left[
  R^{[3,3]}_{j}\left(\cM^{(3,y)}_I ; \mu^{(3)}_I\right)
    J_1^{\rm sub}(m_{j,I}, \vp)\right]\nonumber \\ &&{}
  + a^2\delta'_V\sum_{j\in\cM^{(4,y)}}
  \frac{\partial}{\partial m_{Y,V}^2}\left[
  R^{[4,3]}_{j} \left(\cM^{(4,y)}_V ; \mu^{(3)}_V\right)
    J_1^{\rm sub}(m_{j,V}, \vp)\right]
    \nonumber \\ &&{}  + [V\to A]
  \Biggr\}\ , \eqn{fpD} \\
  %%%
  \left(\delta f_p^{\ref{fig:1loopP}(b)}
  \right)^{B_x\to P_{xy}}_{1\!+\!1\!+\!1}
  &=& \frac{ g^2_\pi}{(4\pi f)^2}
  \Bigg\{
    -\frac{1}{3}\sum_{j\in \cM^{(4,xy)}}
    R_{j}^{[4,3]}\left(\cM^{(4,xy)}_I ; \mu^{(3)}_I\right)
     J_1^{\rm sub}(m_{j,I},\vp)\nonumber \\     && 
    +a^2\delta'_V \sum_{j\in \cM^{(5,xy)}}
    R_{j}^{[5,3]}\left(\cM^{(5,xy)}_V ; \mu^{(3)}_V\right)
    J_1^{\rm sub}(m_{j,V},\vp)
    +\left[ V\to A\right]  \Bigg\}, \eqn{fp5b} \\
    %%%
 	\left( \delta f_p^{\ref{fig:1loopP}(c)}
	\right)^{B_x\to P_{xy}}_{1\!+\!1\!+\!1}
  &=&- \frac{1}{6 (4 \pi f)^2} \Bigg\{
	\frac{1}{16}\sum_{f,\Xi}
 \left[   I_1(m_{xf,\Xi})+I_1(m_{yf,\Xi})\right]
	\nonumber \\ &&{}
	+ \frac{1}{3}\biggl[\sum_{j\in \cM^{(3,y)}}
	\frac{\partial}{\partial m_{Y,I}^2}\left[
	R^{[3,3]}_{j} \left(\cM^{(3,y)}_I ; \mu^{(3)}_I\right)
  I_1(m_{j,I}) \right]\nonumber \\
    && {}
	+ \sum_{j\in \cM^{(3,x)}}
	\frac{\partial}{\partial m_{X,I}^2}\left[
	R^{[3,3]}_{j} \left(\cM^{(3,x)}_I ; \mu^{(3)}_I\right)
  I_1(m_{j,I}) \right]
	\nonumber \\ &&{}
	+2\sum_{j\in \cM^{(4,xy)}}
	\left[
	R^{[4,3]}_{j}\left(\cM^{(4,xy)}_I ; \mu^{(3)}_I\right)
	I_1(m_{j,I}) \right] \biggr]
		\nonumber \\ &&{}
	+ a^2\delta'_V\biggl[\sum_{j\in \cM^{(4,y)}}
	\frac{\partial}{\partial m_{Y,V}^2}\left[
	R^{[4,3]}_{j} \left(\cM^{(4,y)}_V ; \mu^{(3)}_V\right)
  I_1(m_{j,V}) \right]
	\nonumber \\ &&{}+ \sum_{j\in \cM^{(4,x)}}
	\frac{\partial}{\partial m_{X,V}^2}\left[
	R^{[4,3]}_{j} \left(\cM^{(4,x)}_V ; \mu^{(3)}_V\right)
  I_1(m_{j,V}) \right]\nonumber \\ &&{}
	-2\sum_{j\in \cM^{(5,xy)}}
	\left[
	R^{[5,3]}_{j}\left(\cM^{(5,xy)}_V ; \mu^{(3)}_V\right)
	 I_1(m_{j,V}) \right]\biggr]
		+[V\to A] \Bigg\}\ ,
		\eqn{fp5c} \\
		%%%
  \left(\delta f_p^{\ref{fig:1loopP}(d)}
  \right)^{B_x\to P_{xy}}_{1\!+\!1\!+\!1}
  &=& -\frac{1}{2 (4 \pi f)^2} \Bigg\{
    \frac{1}{16}\sum_{f,\Xi}
    I_1(m_{yf,\Xi})\nonumber \\ &&
    +\frac{1}{3}\sum_{j\in \cM^{(3,y)}}
    \frac{\partial}{\partial m_{Y,I}^2}\left[
    R^{[3,3]}_{j}\left(\cM^{(3,y)}_I ; \mu^{(3)}_I\right)
      I_1(m_{j,I})
      \right]\nonumber \\ &&
    \hspace{-0.2cm}+ a^2\delta'_V\sum_{j\in \cM^{(4,y)}}
    \frac{\partial}{\partial m_{Y,V}^2}\left[
    R^{[4,3]}_{j}\left(\cM^{(4,y)}_V ; \mu^{(3)}_V\right)
      I_1(m_{j,V})
      \right] + [V\to A]
    \Bigg\}.
  \eqn{fp5d}
\end{eqnarray}
In \eqsthru{fv4a}{fp5d},
the explicit factors of $1/3$ in front of terms involving the taste-singlet ($I$)
mesons come from the  factors of $1/N_{\rm sea}$ in Ref.~\cite{BECIREVIC}.

To get the full corrections for both $f_v$ and $f_p$, we need to add in the
wavefunction renormalizations, given in Appendix~\ref{app:wf_ren} in
\eqs{deltfv}{deltfp}. Putting these together with the analytic terms and
(for $f_p$) the $D$ term, \eqs{fv}{fp} give the complete NLO expressions
for the form factors in \schpt.

The above $1\!+\!1\!+\!1$ results are expressed in terms of the Euclidean residue functions
 $R^{[n,k]}_j$, \eq{residues}. In the $2\!+\!1$ case, there is a cancellation in the
residues between the
contribution of the $U$ or $D$ in the numerator and that of the $\pi_0$ in the
denominator. Thus,
to obtain the $2\!+\!1$ from the $1\!+\!1\!+\!1$ case, one must simply reduce by one all
superscripts on the residues, \ie $R^{[n,k]}\to R^{[n-1,k-1]}$, and remove $m_{\pi_0}$
and (say) $m_D$ from the mass sets:
\begin{eqnarray}
  \mu^{(3)} & \to & \{m^2_U,m^2_S\}\label{eq:num_masses2}\ ,\\*
  \cM^{(3,x)} & \to & \{m_X^2, m_{\eta}^2\}
  \label{eq:den_masses_xI2}\ ,\\*
  \cM^{(4,x)} & \to & \{m_X^2, m_{\eta}^2, m_{\eta'}^2\}
  \label{eq:den_masses_xA2}\ , \\*
  \cM^{(4,xy)} & \to & \{m_X^2,m_Y^2, m_{\eta}^2\}
  \label{eq:den_masses_xyI2}\ ,\\*
  \cM^{(5,xy)} & \to & \{m_X^2,m_Y^2,
  m_{\eta}^2, m_{\eta'}^2\}
  \label{eq:den_masses_xyA2}\ .
\end{eqnarray}

We also write here the expressions for three
non-degenerate dynamical flavors in continuum PQ\chpt, which to our knowledge
do not appear in the literature.
These expressions can be obtained either by returning to
Ref.~\cite{BECIREVIC} and using the residue functions to
generalize to the non-degenerate case, or simply by taking
the continuum limit of the above equations. Either way, the results for
$f_v$ are
\begin{eqnarray}
  \left(\delta f_v^{\ref{fig:1loopV}(a),\rm cont}
  \right)^{B_x\to P_{xy}}
  & = & \frac{1}{2 (4 \pi f)^2} \Biggl\{
	\sum_{f}
  \Bigl[I_1(m_{yf}) + 2 I_2(m_{yf},\vp)\Bigr]
  \nonumber\\&&{}
  +\frac{1}{3}\Biggl[\sum_{j\in \cM^{(4,xy)}}
  R^{[4,3]}_{j}\left(\cM^{(4,xy)} ; \mu^{(3)}\right)
    \left[I_1(m_{j}) + 2I_2(m_{j},\vp)\right]
    \nonumber\\&&+ \frac{\partial}{\partial m_{Y}^2}\bigg(
    \sum_{j\in \cM^{(3,y)}}
    R^{[3,3]}_{j}\left(\cM^{(3,y)} ; \mu^{(3)}\right)
    \left[I_1(m_{j}) + 2I_2(m_{j},\vp)\right]\bigg)
    \Biggr] \Biggr\} \ ,
  \nonumber \\
%%%%%%%%%%%%%%%%%%%%%%%%%%%%%%%%%%%%
 \left( \delta f_v^{\ref{fig:1loopV}(b),\rm cont}
  \right)^{B_x\to P_{xy}}
  &=&-\frac{1}{6 (4 \pi f)^2}\Biggl\{
  \sum_{f} \left[I_1(m_{xf}) + I_1(m_{yf}) \right]
  \nonumber\\&&{}
  +\frac{1}{3}\biggl[ \frac{\partial}{\partial m_{Y}^2}\bigg(
    \sum_{j\in \cM^{(3,y)}}
    R^{[3,3]}_{j}\left(\cM^{(3,y)} ; \mu^{(3)}\right)
    I_1(m_{j}) \bigg) \nonumber\\&&{}
    + \frac{\partial}{\partial m_{X}^2}\bigg(
    \sum_{j\in \cM^{(3,x)}}
    R^{[3,3]}_{j} \left(\cM^{(3,x)} ; \mu^{(3)}\right)
    I_1(m_{j}) \bigg) \nonumber\\&&{}
    -\sum_{j\in \cM^{(4,xy)}}
    R^{[4,3]}_{j}\left(\cM^{(4,xy)} ; \mu^{(3)}\right)
    I_1(m_{j})
     \biggr]
  \Biggr\}  \ , \eqn{fvPQcont}
\end{eqnarray}
while those for $f_p$ are
\begin{eqnarray}
  \left(D^{\rm cont} \right)^{B_x\to P_{xy}}
  &=& - \frac{3g_\pi^2\vp}{(4 \pi f)^2}
  \Biggl\{ \sum_{f}
  J_1^{\rm sub}(m_{yf}, \vp)\nonumber \\ &&{}
  + \frac{1}{3}\sum_{j\in \cM^{(3,y)}}
  \frac{\partial}{\partial m_{Y}^2}\left[
  R^{[3,3]}_{j}\left(\cM^{(3,y)}; \mu^{(3)}\right)
    J_1^{\rm sub}(m_{j}, \vp)\right]\Biggr\}\ , \nonumber \\
%%%%%%%%%%%%%%%%%%%%%%%%%%%%%%%%%
  \left(\delta f_p^{\ref{fig:1loopP}(b),\rm cont}
  \right)^{B_x\to P_{xy}}
  &=& \frac{ g^2_\pi}{(4\pi f)^2}
  \Bigg\{
    -\frac{1}{3}\sum_{j\in \cM^{(4,xy)}}
    R_{j}^{[4,3]}\left(\cM^{(4,xy)} ; \mu^{(3)}\right)
     J_1^{\rm sub}(m_{j},\vp)\Bigg\}\ , \nonumber \\
%%%%%%%%%%%%%%%%%%%%%%%%%%%%%%%%%
 	\left( \delta f_p^{\ref{fig:1loopP}(c),\rm cont}
	\right)^{B_x\to P_{xy}}
  &=&- \frac{1}{6 (4 \pi f)^2} \biggl\{
	\sum_{f}\left[I_1(m_{xf})+I_1(m_{yf})\right]
	\nonumber \\ &&{}
	+ \frac{1}{3}\biggl[\sum_{j\in \cM^{(3,y)}}
	\frac{\partial}{\partial m_{Y}^2}\left[
	R^{[3,3]}_{j} \left(\cM^{(3,y)} ; \mu^{(3)}\right)
  	I_1(m_{j}) \right]\nonumber \\
    && {}
	+ \sum_{j\in \cM^{(3,x)}}
	\frac{\partial}{\partial m_{X}^2}\left[
	R^{[3,3]}_{j} \left(\cM^{(3,x)} ; \mu^{(3)}\right)
  	I_1(m_{j}) \right]
	\nonumber \\ &&{}
	+2\sum_{j\in \cM^{(4,xy)}}\left[
	R^{[4,3]}_{j}\left(\cM^{(4,xy)} ; \mu^{(3)}\right)
	I_1(m_{j}) \right] \biggr] \biggr\}\ ,
		\nonumber \\
%%%%%%%%%%%%%%%%%%%%%%%%%%%%%%%%%
  \left(\delta f_p^{\ref{fig:1loopP}(d),\rm cont}
  \right)^{B_x\to P_{xy}}
  &=& -\frac{1}{2 (4 \pi f)^2} \Bigg\{
    \sum_{f}I_1(m_{yf})\nonumber\\&&{}
    +\frac{1}{3}\sum_{j\in \cM^{(3,y)}}
    \frac{\partial}{\partial m_{Y}^2}\left[
    R^{[3,3]}_{j}\left(\cM^{(3,y)} ; \mu^{(3)}\right)
      I_1(m_{j})
      \right]\Bigg\}\, .
  \label{eq:fpPQcont}
\end{eqnarray}
Corresponding continuum-limit results for the wave-function renormalizations
are given in Appendix \ref{app:wf_ren}.

\subsection{Full QCD Results}\label{sec:fullQCD}

Adding together the complete results for the ``full QCD'' case is
straightforward. For simplicity, we specialize to case $m_u=m_d$ (\ie $2+1$). 
For $B\to\pi$, the complete corrections (including wave-function renormalization) are:
\begin{eqnarray}
	D^{B\to\pi} & = &  - \frac{3g_\pi^2\vp}{(4 \pi f)^2}
  \Biggl\{ \frac{1}{16}\sum_{\Xi}
  \left[2J_1^{\rm sub}(m_{\pi,\Xi}, \vp )+
  J_1^{\rm sub}(m_{K,\Xi}, \vp) \right]
  \nonumber \\ &&{}
  -\frac{1}{2}J_1^{\rm sub}(m_{\pi,I}, \vp)
 	+\frac{1}{6}J_1^{\rm sub}(m_{\eta,I}, \vp)
    \nonumber \\ &&{}+
  	\sum_{j\in\{\pi,\eta,\eta'\}}
	\left[(- a^2\delta'_V)
  	R^{[3,1]}_{j} \left(\{m_{\pi,V},m_{\eta,V},m_{\eta',V}\} ;
	 \{m_{S,V}\}\right)
    J_1^{\rm sub}(m_{j,V}, \vp)\right] \nonumber\\
    &&{}+ \bigl[V\to A\bigr]
  \Biggr\} \ , \eqn{B-to-pi-D}
\end{eqnarray}
\begin{eqnarray}
  \delta f_p^{B\to\pi} & = &
  \frac{1}{(4 \pi f)^2}\Biggl\{ 	  \frac{1}{16}\sum_{\Xi}\left[
  -  \frac{1+3 g_\pi^2}{2}\left[ 2I_1(m_{\pi,\Xi})
  + I_1(m_{K,\Xi})\right]
  \right]\nonumber\\
  &&
   -\frac{1}{2}g^2_\pi J_1^{\rm sub}(m_{\pi,I}, \vp)
   +\frac{1}{6}g^2_\pi J_1^{\rm sub}(m_{\eta,I}, \vp)
  +\frac{1+3 g^2_\pi}{12}
  \biggl[3I_1(m_{\pi,I}) - I_1(m_{\eta,I}) \biggr]
  \nonumber\\&&
  {}+
  \sum_{j\in\{\pi,\eta,\eta'\}}
  \biggl[ a^2\delta'_V
    R^{[3,1]}_{j} \left(\{m_{\pi,V},m_{\eta,V},m_{\eta',V}\} ;
	 \{m_{S,V}\}\right)\nonumber\\&&{}\times
    \left(
    g_\pi^2 J_1^{\rm sub}(m_{j,V}, \vp)
  	+\frac{1+3g^2_\pi}{2}
    I_1(m_{j,V})\right) \biggr] + [V\to A]\Biggr\}\ , \eqn{B-to-pi-fp}
\end{eqnarray}
\begin{eqnarray}
  \delta f_v^{B\to\pi} & = &
  \frac{1}{(4 \pi f)^2}\Biggl\{ 	  \frac{1}{16}\sum_{\Xi}\biggl[
  \frac{1-3g^2_\pi}{2}\left[
    2I_1(m_{\pi,\Xi}) + I_1(m_{K,\Xi})
    \right]\hspace{5truecm}\nonumber\\&&{}
  +2I_2(m_{\pi,\Xi},\vp) + I_2(m_{K,\Xi},\vp)
    \biggr]\nonumber\\
	& & {}+\frac{1+3g^2_\pi}{4}\left[ I_1(m_{\pi,I})
  - \frac{1}{3}I_1(m_{\eta,I})\right]\nonumber\\
 	& & {}+
  \sum_{j\in\{\pi,\eta,\eta'\}}
   \biggl[a^2\delta'_V
    R^{[3,1]}_{j} \left(\{m_{\pi,V},m_{\eta,V},m_{\eta',V}\} ;
	 \{m_{S,V}\}\right)\nonumber\\&&{}\times
  \left(\frac{3(g^2_\pi-1)}{2}
	 I_1(m_{j,V})-2  I_2(m_{j,V},\vp)
	 \right)\biggr] + [V\to A]\Biggr\}\ . \eqn{B-to-pi-fv}
\end{eqnarray}

For $B\to K$,\footnote{The transition $B\to K$ occurs through
penguin diagrams; $D\to K$ is a standard semileptonic decay due to the
current in \eq{LOcurrent}. We keep the notation $B\to K$ however to stress
that we are working to lowest order in the heavy quark mass.} we have
\begin{eqnarray}
	D^{B\to K} & = &  - \frac{3g_\pi^2(\vp)}{(4 \pi f)^2}
  	\Biggl\{ \frac{1}{16}\sum_{\Xi}
  	\left[2J_1^{\rm sub}(m_{K,\Xi}, \vp )+
  	J_1^{\rm sub}(m_{S,\Xi}, \vp) \right]
  	\nonumber \\ &&{}
  	+\frac{2}{3}J_1^{\rm sub}(m_{\eta,I}, \vp)
 	-J_1^{\rm sub}(m_{S,I}, \vp)
    \nonumber \\ &&{}+
  	\sum_{j\in\{S,\eta,\eta'\}}
	\left[(- a^2\delta'_V)
  	R^{[3,1]}_{j} \left(\{m_{S,V},m_{\eta,V},m_{\eta',V}\} ;
	 \{m_{\pi,V}\}\right)
    J_1^{\rm sub}(m_{j,V}, \vp)\right] \nonumber\\
    &&{}+ \bigl[V\to A\bigr] \Biggr\} \ ,\eqn{B-to-K-D}
\end{eqnarray}
\begin{eqnarray}
  	\delta f_p^{B\to K} & = &
  	\frac{1}{(4 \pi f)^2}\Biggl\{ 	  	\frac{1}{16}\sum_{\Xi}\left[
  	-  \frac{2+3 g_\pi^2}{2} I_1(m_{K,\Xi})
  	-\frac{1}{2} I_1(m_{S,\Xi})
  	-3g_\pi^2 I_1(m_{\pi,\Xi})
  	\right]\hspace{1truecm}\nonumber\\&&
   	-\frac{1}{3}g^2_\pi J_1^{\rm sub}(m_{\eta,I}, \vp)
  	+\frac{3 g^2_\pi}{4}I_1(m_{\pi,I})
	- \frac{4+3 g^2_\pi}{12}I_1(m_{\eta,I})
	+ \frac{1}{2}I_1(m_{S,I})
  	\nonumber\\&&{}
	+ a^2\delta'_V	\biggl[
	\frac{g_\pi^2}{m^2_{\eta',V}-m^2_{\eta,V}}
	\biggl(J_1^{\rm sub}(m_{\eta,V}, \vp)
	-J_1^{\rm sub}(m_{\eta',V}, \vp)
	\biggr)\nonumber\\&&{}
  	+\frac{3g^2_\pi}{2}\sum_{j\in\{\pi,\eta,\eta'\}}
    	R^{[3,1]}_{j} \left(\{m_{\pi,V},m_{\eta,V},m_{\eta',V}\} ;
	\{m_{S,V}\}\right) I_1(m_{j,V})
	\nonumber\\&&{}
  	+\frac{1}{2}\sum_{j\in\{S,\eta,\eta'\}}
    	R^{[3,1]}_{j} \left(\{m_{S,V},m_{\eta,V},m_{\eta',V}\} ;
	\{m_{\pi,V}\}\right) I_1(m_{j,V})
     \biggr] \nonumber \\
&&{} 
+ [V\to A]\Biggr\}\ , \eqn{B-to-K-fp}
%%%
\end{eqnarray}
\begin{eqnarray}
  	\delta f_v^{B\to K} & = &
  	\frac{1}{(4 \pi f)^2}\Biggl\{
\frac{1}{16}\sum_{\Xi}\biggl[
      \frac{2-3g^2_\pi}{2} I_1(m_{K,\Xi})
    - 3g^2_\pi I_1(m_{\pi,\Xi})
    + \frac{1}{2}I_1(m_{S,\Xi})\nonumber\\&&{}
  	+2I_2(m_{K,\Xi},\vp) + I_2(m_{S,\Xi},\vp)
    \biggr]
    \nonumber\\& & {}
	- \frac{1}{2} I_1(m_{S,I})
  	+ \frac{3 g^2_\pi}{4} I_1(m_{\pi,I})
	+ \frac{8-3g^2_\pi}{12} I_1(m_{\eta,I})
  	+ I_2(m_{\eta,I},\vp) - I_2(m_{S,I},\vp)
	\nonumber\\ & & {}
	+ a^2\delta'_V \Biggl[
	\frac{I_1(m_{\eta',V}) - I_1(m_{\eta,V})
	+I_2(m_{\eta',V},\vp) - I_2(m_{\eta,V},\vp)}
	{m^2_{\eta',V} - m^2_{\eta,V}}
	\nonumber\\&&{}
	- \sum_{j\in\{S,\eta,\eta'\}}
    R^{[3,1]}_{j} \left(\{m_{S,V},m_{\eta,V},m_{\eta',V}\} ;
	\{m_{\pi,V}\}\right)\left(
	\frac{1}{2}I_1(m_{j,V}) + I_2(m_{j,V}, \vp)
	\right)\nonumber\\&&{}
	+ \frac{3g^2_\pi}{2}\sum_{j\in\{\pi,\eta,\eta'\}}
    R^{[3,1]}_{j} \left(\{m_{\pi,V},m_{\eta,V},m_{\eta',V}\} ;
	\{m_{S,V}\}\right) I_1(m_{j,V}) \Biggr]
	\nonumber\\&&{}+ [V\to A]\Biggr\}\ . \eqn{B-to-K-fv}
\end{eqnarray}

\subsection{Analytic terms}\label{sec:analytic}
  From the power counting discussed 
in \secref{hqetschpt},  as well as interchange symmetry among
the sea quark masses, 
the form factors at the order we are working
can only depend only on the valence quark masses $m_x$ and $m_y$,
the sum of the sea quark masses $m_u + m_d + m_s$, the pion momentum (through $\vp$),
and the lattice spacing, $a$.  
The last must appear quadratically, since the errors of the staggered action are $\cO(a^2)$.
Recall that we do not include any discretization errors
coming from the heavy quark in our effective theory.

Thus we expect to have 
the analytic terms shown in \eqs{fv}{fp} with coefficients $c^p_i$ and  $c^v_i$.  
(Here $i=\{x,y,{\rm sea}, 1,2,a\}$.)  
We then can examine, one by one, the known NLO terms in the Lagrangian and current
to check for the existence of relations among the $c^p_i$ and/or $c^v_i$.  As soon as
a sufficient number of terms are checked to ensure that the parameters are
independent, we are done.  It is therefore  not necessary in all cases to
have a complete catalog of NLO terms. Unless otherwise indicated, all NLO terms discussed
in this section come from Ref.~\cite{HL_SCHPT}.

Note first of all that we do not need to include explicitly the effects of mass-renormalization
terms in the NLO heavy-light Lagrangian, such as
\begin{equation}\label{eq:L2m}
   2\lambda_1 \Tr\left(\overline{H} H\cM^+\right) + 2\lambda'_1 \Tr\left(\overline{H} H\right)\Tr\left(\cM^+\right)   \ ,
\end{equation}
where we define
\begin{equation}\label{eq:Mpm}
  \cM^\pm  =  \frac{1}{2}\left(\sigma \cM\sigma
  \pm \sigma^{\dagger} \cM\sigma^{\dagger}\right) \ .
\end{equation}
The effect of the terms in \eq{L2m} is absorbed into the $B^*_y$-$B_x$ mass difference $\Delta^*_{yx}$,
\eq{Deltaxy}, just like the one-loop contribution to the mass.  Corresponding $\cO(a^2)$
term in the Lagrangian, which can be obtained by replacing $\cM^+$ above
by various taste-violating operators, can likewise be ignored here.

We now consider the discretization corrections parametrized by $c^p_a$ and $c^v_a$.  
There are a large
number of $\cO(a^2)$ corrections to the Lagrangian and the current that can contribute
to these coefficients, so it is
not surprising that they are independent.  For example, consider the
following terms in the NLO heavy-light Lagrangian 
\begin{equation}\eqn{fp-a2}
a^2\sum_{k=1}^8 c^A_{3,k} \Tr\left( \overline{H}
     H \gamma_\mu\gamma_5\{\mathbb{A}^\mu,
    \cO^{A,+}_k\}\right)\ ,
\end{equation}
where the $\cO^{A,+}_k$ are various taste-violating operators, similar to those in \eq{VSigma} above.
These terms do not contribute to $c^v_a$, but only to $c^p_a$, though 
corrections to the $B$-$B^*$-$\pi$ vertex in \figref{tree}(b).  On the other hand, there
are many terms that contribute both to  $c^v_a$ and to $c^p_a$.  An example is the
following correction to the current 
\begin{equation}\eqn{j-a2}
a^2\sum_{k=1}^8
  r^A_{1,k}\,
  \trD\left(\gamma^\mu (1-\gamma_5) H  \right)
  \cO^{A,+}_k \sigma^\dagger \lambda^{(b)}\ ,
\end{equation}
which contributes equally to $c^v_a$ and $c^p_a$.   Additional examples are provided by those
terms with two derivatives
in the $\cO(m_q a^2)$ pion Lagrangian \cite{Sharpe:2004is}, which correct both coefficients
though their effect on the pion wave-function renormalization.

We consider the $\vp$ and $(\vp)^2$ terms next, namely $c^v_1$, $c^v_2$, $c^p_1$, and $c^p_2$.
This is a case where a complete catalog of Lagrangian and current corrections does not exist.
However, it is easy to find corrections that contribute only to $f_v$ or only to
$f_p$.  As in the previous case, corrections to the $B$-$B^*$-$\pi$ vertex in \figref{tree}(b)
only affect $f_p$ at the order we are working.  Thus,
\begin{equation}\eqn{L2k}
\frac{i\epsilon_1}{\Lambda_\chi}\Tr\left((v\cdot \rightvec D\, \overline{H}H
- \overline{H}H v\cdot\leftvec D)\,\gamma_\mu \gamma_5 \mathbb{A}^\mu\right)
\end{equation}
contributes to $c^p_1$ only; while
\begin{equation}\eqn{L3k}
\frac{\epsilon_3}{\Lambda_\chi^2}\Tr\left(\overline{H}H \gamma_\mu \gamma_5
(v\cdot \rightvec D\,)^2 \mathbb{A}^\mu\right)
\end{equation}
contributes to $c^p_2$ only.  
Similarly, only $f_v$ is affected, though \figref{tree}(a),
by any correction to the current whose expansion in terms of
pion fields starts at linear order (\ie corrections of schematic form 
$H (\frac{i\Phi}{2f} + \cdots)$, with $\cdots$ denoting higher order
terms in $\Phi$).
Thus,
\begin{equation}\eqn{j1k}
\frac{\kappa_2}{\Lambda_\chi}\;
  \trD\bigl(\gamma^\mu \left(1\!-\!\gamma_5\right) H\,\bigr)
v\cdot  \mathbb{A}\, \sigma^\dagger \lambda^{(b)} 
\end{equation}
contributes to $c^v_1$ only; while
\begin{equation}\eqn{j2k}
\frac{i\kappa_4}{\Lambda^2_\chi}\;
  \trD\bigl(\gamma^\mu \left(1\!-\!\gamma_5\right) H\,\bigr)
v\cdot \rightvec D\, v\cdot  \mathbb{A}\, \sigma^\dagger \lambda^{(b)}
\end{equation}
contributes to $c^v_2$ only.  
Since there is at least one Lagrangian or current term that
contributes to each of $c^v_1$, $c^v_2$, $c^p_1$, and $c^p_2$ exclusively, these
coefficients are independent.

The argument for the independence of the sea-quark mass terms, \ie the coefficients $c^v_{\rm sea}$
and $c^p_{\rm sea}$, is similar.  
The  Lagrangian correction 
\begin{equation}\eqn{L3m-k4}
 k_4 \Tr\left( \overline{H}
     H \gamma_\mu\gamma_5 \mathbb{A}^\mu \right)
     \Tr(\cM^+)
\end{equation}
contributes to $c^p_{\rm sea}$ only; while the current correction 
\begin{equation}\eqn{j2m-rho2}
\rho_2\,
  \trD\left(\gamma^\mu (1-\gamma_5) H\right) \sigma^\dagger
   \lambda^{(b)}
  \Tr(\cM^+) 
\end{equation}
contributes equally to both $c^p_{\rm sea}$ and $c^v_{\rm sea}$.  These two observations
are enough to guarantee that $c^v_{\rm sea}$
and $c^p_{\rm sea}$ are independent.

We now turn to the coefficients that control the valence quark mass dependence of the form
factors: $c^v_x$, $c^v_y$,  $c^p_x$, and $c^p_y$. At first glance, it would seem unlikely
that there could be any constraint among these parameters, since there are seven
terms in the Lagrangian and current in Ref.~\cite{HL_SCHPT} that could generate valence mass
dependence.\footnote{There are additional terms involving $ \Tr(\cM^+)$, as in \eqs{L3m-k4}{j2m-rho2},
that only give sea quark mass dependence at this order.} However, three of these terms are immediately eliminated, either because
they could only contribute to flavor-neutral pions (with $x=y$), or because they produce
no fewer than two pions.  There are then two remaining corrections to the heavy-light
Lagrangian,
\begin{equation}
i k_1 \Tr\left( \overline{H}H  v\negcdot \leftvec D\, \cM^+ - v\negcdot \rightvec D \,
\overline{H}H\, \cM^+
  \right) 
  +
  k_3 \Tr\left( \overline{H}
     H  \gamma_\mu\gamma_5\{\mathbb{A}^\mu ,
  \cM^+\}\right)\eqn{L3m-k1-k3}
\end{equation}
and two corrections to the current,
\begin{equation}\eqn{j2m-rho1-rho3}
 \rho_1\,
  \trD\left(\gamma^\mu (1-\gamma_5) H  \right)
  \cM^+ \sigma^\dagger \lambda^{(b)}
+ \rho_3\, \trD\left(\gamma^\mu (1-\gamma_5) H\right)
\cM^- \sigma^\dagger \lambda^{(b)}
\end{equation}

The $k_3$ term in  \eq{L3m-k1-k3} contributes only to $f_p$, through the 
 $B$-$B^*$-$\pi$ vertex.  However, because of the anticommutator,
its contribution is proportional to $m_x+m_y$, so it gives equal contributions
to $c^p_x$ and $c^p_y$. Similarly, because the one-pion term in $\cM^-$ is proportional
to $\Phi \cM + \cM \Phi$, the $\rho_3$ term contributes equally to  $c^v_x$ and $c^v_y$
(but not at all to  $c^p_x$ and $c^p_y$).  Further, since $\cM^+$ creates
only even number of pions, we can replace it by $\cM$ in \eq{j2m-rho1-rho3}.
The $\rho_1$ term can then easily be seen to contribute equally to $c^p_y$ and $c^v_x$,
since the current needs to annihilate a $B^*_y$ in the $f_p$ case and a
$B_x$ in the $f_v$ case.

The contributions of the $k_1$ term in  \eq{L3m-k1-k3} are the most non-trivial.
It contributes to both $c^v_x$ and $c^p_x$ through wave function renormalization
on the external $B_x$ line,  but it also contributes to $c^p_y$ through an insertion
on the internal $B^*_y$ line in \figref{tree}(b).  However, since wave-function renormalization
effects on external lines go like $\sqrt{Z}$, 
the contributions of this term to both  $c^v_x$ and $c^p_x$ are exactly half
of its contribution to  $c^p_y$.  Thus, all four terms in \eqs{L3m-k1-k3}{j2m-rho1-rho3}
are consistent with the relation given in \eq{relation}.

We still need to worry about valence mass dependence generated by the standard 
$\cO(p^4)$ pion Lagrangian \cite{GASSER} through wave function renormalization
of the external pion. Such contributions do exist (from $L_5$), but the $x\leftrightarrow y$ symmetry
of the pion guarantees they are proportional to $m_x+m_y$ in both $f_p$ and $f_v$,
and hence do not violate \eq{relation}.

A consistency check of the relation, \eq{relation}, as well as of the claimed independence of the
other analytic terms, can be performed by considering the change in the chiral logarithms in
\eqsthru{fv4a}{fp5d} and \eqs{ZP}{ZB} under a change in chiral scale.  To simplify 
the calculation, it is very convenient to use the conditions obeyed by sums of residues,
which are given in Eq.~(38) of the second paper in Ref.~\cite{SCHPT}.
We find that such a scale change can be absorbed by parameters that obey \eq{relation}
but are otherwise independent.  

In the continuum limit, 
$c^p_{\rm sea}$ and $c^v_{\rm sea}$   remain independent,
as do $c^p_1$,  $c^p_2$,  $c^v_1$,  and $c^v_2$.
We disagree on these points with Ref.~\cite{BECIREVIC},
which found  $c^p_{\rm sea}= c^v_{\rm sea}$, and did not consider analytic terms giving $\vp$
dependence.  The difference can be traced to the inclusion here of the effects of the complete
set of NLO mass-dependent terms, as well as a sufficient number of higher derivative
terms (\eqsthru{L2k}{j2k}).  In particular, the independence of $c^p_{\rm sea}$ and $c^v_{\rm sea}$
can be traced to the existence of the Lagrangian correction, \eq{L3m-k4}, which was not
considered in  Ref.~\cite{BECIREVIC}.  On the other hand, the relation among the valence
mass coefficients, \eq{relation}, is obeyed by the expressions for these coefficients found
in  Ref.~\cite{BECIREVIC}.  This occurs because the contributions of the terms proportional
to $k_3$ and $\rho_3$ in \eqs{L3m-k1-k3}{j2m-rho1-rho3}, which were not considered in  Ref.~\cite{BECIREVIC}, are proportional to $m_x+m_y$ and automatically obey \eq{relation}.

Note, finally, that the relation in \eq{relation} is almost certain to be violated
at next order in HQET. This is because the contributions from operators
like the $k_1$ term in \eq{L3m-k1-k3} will affect the $B$ and the $B^*$ differently
at $\cO(1/m_Q)$, destroying the cancellation that made \eq{relation} possible.  

\section{Finite Volume Effects}\label{sec:FV}

In a finite volume, we must replace the integrals in
\eqsthru{I1int}{J1J2int} by discrete momentum sums. 
We assume that the time direction is large enough to be considered
infinite (this is the case in MILC simulations), and that each of the spatial lengths has 
(dimensionful) size $L$.

The correction to \eq{I1int} is given explicitly in Ref.~\cite{CHIRAL_FSB}. In
finite volume, we need
only make the replacement
\begin{equation}\label{eq:I1_FV}
	I_1(m) \to I^{\rm fv}_1(m) = I_1(m) + m^2 \delta_1(mL)\ .
\end{equation}
Here $\delta_1$ is a sum over modified Bessel functions
\begin{equation}\label{eq:delta1}
	\delta_1(mL) = \frac{4}{mL}
    \sum_{\vec r\ne 0}
    \frac{K_1(rmL)}{r}\ ,
\end{equation}
where $\vec r$ is a 3-vector with integer components, and $r\equiv |\vec r\,|$.

Arndt and Lin \cite{LIN_ARNDT} have worked out the finite
volume correction to \eq{I2int}.  In our notation, the function $ I_2(m,\Delta)$ is replaced
by its finite volume form, $ I^{\rm fv}_2(m,\Delta)$, 
\begin{equation}\label{eq:I2_FV}
	I_2(m,\Delta) \to I^{\rm fv}_2(m,\Delta)= I_2(m,\Delta) + \delta I_2(m,\Delta,L)
	 \ ,
\end{equation}
where the correction $\delta I_2(m,\Delta,L)$ is given simply in terms of the function
$J_{\rm FV}(m,\Delta,L)$ defined in Eq.~(44) of Ref.~\cite{LIN_ARNDT}:\footnote{We have added
the $L$ argument to $J_{\rm FV}$ for consistency with our notation}
\begin{eqnarray}
	\delta I_2(m,\Delta,L)  &=& -(4\pi)^2\, \Delta\; J_{\rm FV}(m,\Delta,L) \nonumber \\
	 J_{\rm FV}(m,\Delta,L) &\equiv& \left(\frac{1}{2\pi}\right)^2 \sum_{\vec r \ne 0}
	\int_0^\infty dq
	\left(\frac{q}{\omega_q (\omega_q + \Delta)}\right)
	\left(\frac{\sin(qr L)}{rL}\right)\ , \eqn{JFV-defn}
\end{eqnarray}
with $\omega_q = \sqrt{q^2 + m^2}$.

The asymptotic form of $J_{\rm FV}(m,\Delta,L)$ for large $mL$ is useful for practical
applications, where typically $mL>3$, and often $mL>4$ \cite{SHIGEMITSU,Aubin:2004ej}.
Arndt and Lin have found \cite{LIN_ARNDT}:
\begin{eqnarray}
J_{\rm FV}(m,\Delta,L) 
	& = &
	\sum_{\vec r \ne 0}
	\left(\frac{1}{8 \pi r L}\right)
	e^{-rmL}\cA
	\label{eq:JFV-exp}\ , \\
	\cA & = &
	e^{(z^2)} \left[ 1 - {\mathrm{Erf}}(z)\right]
	+\left (\frac{1}{rm L} \right ) \bigg [
 	\frac{1}{\sqrt{\pi}} \left ( \frac{z}{4} -
	\frac{z^{3}}{2}\right )
 	+ \frac{z^{4}}{2}e^{(z^2)}
 	\big [ 1 - {\mathrm{Erf}}(z)\mbox{ }\big ]
	\bigg ]\nonumber\\& &
	\hspace{-0.2cm}-\left (\frac{1}{rm L} \right )^2\bigg [
	\frac{1}{\sqrt{\pi}}\left ( \frac{9z}{64} -
	\frac{5z^{3}}{32}
  	+\frac{7z^{5}}{16} + \frac{z^{7}}{8} \right )
	-\left ( \frac{z^{6}}{2} + \frac{z^{8}}{8}\right )
	e^{(z^2)} \big [ 1 - {\mathrm{Erf}}(z)\big ]
	\bigg ]\nonumber\\&&{}
	+\cO\left(\frac{1}{(rm L)^3}\right)
	\label{eq:A}\ ,
\end{eqnarray}
where
\begin{equation}
	z \equiv \left (\frac{\Delta}{m}\right )
	\sqrt{\frac{rmL}{2}} \ .
\end{equation}
Computing higher orders in the $1/(mL)$ expansion is possible if greater
precision is needed.

Since the functions $I_1(m)$ and $I_2(m,\Delta)$ arise from the 
integral $\cI_3^\mu(m,\Delta)$ in
\eq{I1I2int}, as well as from \eqs{I1int}{I2int}, which serve to define them,
it is necessary to check that the finite volume corrections coming from
\eq{I1I2int} are just those given by \eqs{I1_FV}{I2_FV} above.  This is
easily seen to be true in the rest frame of the heavy quark, in which we are
working.  It is a consequence of the facts that: (1) in the rest frame,
only the $\mu=0$ component of $\cI_3^\mu(m,\Delta)$ is non-zero, and (2) the
integral over $dq^0$ is unaffected by finite volume, since we assume
large time-extent of the lattices. The finite volume integral then splits into 
$I^{\rm fv}_1(m)$ and $I^{\rm fv}_2(m,\Delta)$ pieces, just as  in infinite volume.

Finally,  we have to examine the finite volume corrections to the
integral $\cJ^{\mu\nu}$, \eq{J1J2int}.
Since the function 
$J_2(m,\Delta)$ does not 
enter our final results, we need
only evaluate
\begin{eqnarray}\label{eq:PmunuJmunu}
	\cJ \equiv
	(g_{\nu\mu} - v_\nu v_\mu)\cJ^{\mu\nu}
	& = & \int\frac{d^4 q}{(2\pi)^4}
	\frac{i(g_{\nu\mu} - v_\nu v_\mu)q^\mu q^\nu}
	{(\vq - \Delta+i\epsilon)(q^2-m^2+i\epsilon)} \nonumber\\
	& = & \int\frac{d^4 q}{(2\pi)^4}
	\frac{-i\mathbf{q}^2}
	{(\vq - \Delta+i\epsilon)(q^2-m^2+i\epsilon)} \nonumber\\
	& \to &
	\frac{3\Delta}{(4\pi)^2}
	J_1(m,\Delta)\ ,
\end{eqnarray}
where $\mathbf{q}$ is the spatial 3-vector part of $q^\mu$.
In the last line, the arrow refers to the fact that the function $J_1$ arises 
after regularization and renormalization of the integral.
A useful regulator in the present
context is given by the insertion of a factor of $\exp(-\omega_q/\Lambda_0)$, where
$\Lambda_0$ is a cutoff.
After performing the contour integral over $q^0$,
\begin{eqnarray}\label{eq:J_finvol1}
	\cJ
	& = &
	\int\frac{d^3 q}{(2\pi)^3}
	\frac{\mathbf{q}^2}
	{2\omega_q(\omega_q + \Delta)}\nonumber\\
	& = &
	\int\frac{d^3 q}{(2\pi)^3}
	\frac{1}{2}
	-\int\frac{d^3 q}{(2\pi)^3}
	\frac{\Delta}
	{2\omega_q}
	+\int\frac{d^3 q}{(2\pi)^3}
	\frac{\Delta^2-m^2}{2\omega_q(\omega_q + \Delta)}\ .
\end{eqnarray}
The first term is a pure divergence with no $m$ or $\Delta$ dependence. It is thus the same
in finite volume or infinite volume \cite{CHIRAL_FSB}. 
The correction to the middle term is 
proportional to the correction to $I_1$, since the same integral
appears after performing the $q^0$ integration in \eq{I1int}.  
Similarly, the integral in the 
third term is proportional to that arising from the  $q^0$ integration in \eq{I2int},
and the correction is therefore already known.
We have 
\begin{equation}\label{eq:J1_replacement}
	J_1(m,\Delta) \to J^{\rm fv}_1(m,\Delta) = 
	J_1(m,\Delta) + \delta J_1 (m,\Delta,L)\ ,
\end{equation}
where
\begin{equation}\label{eq:deltaJ1_full}
	\delta J_1 (m,\Delta,L)
	= 
	\frac{m^2-\Delta^2}{3\Delta^2} \; \delta I_2(m,\Delta,L)
-\frac{m^2}{3}\;\delta_1(mL) 
\end{equation}
The correction to $J_1^{\rm sub}$, \eq{sub_J1}, is 
\begin{equation}
\delta J_1^{\rm sub} (m,\Delta,L)  =  \delta J_1 (m,\Delta,L) 
+ \frac{16\pi^2 m^2}{3\Delta} J_{\rm FV}(m,0,L) \ ,
\end{equation}
where $J_{\rm FV}(m,0,L)$ is the same as \eq{JFV-exp} with $\mathcal{A}=1$.

With the expressions in this section, it is straightforward to incorporate the corrections
to $I_1$, $I_2$, and $J_1$ numerically into fits to finite-volume lattice data.

\section{Conclusions}\label{sec:conc}

We have presented the NLO expressions in partially quenched \schpt\ 
for the form factors associated with $B\to P_{xy}$ semileptonic decays, 
for both infinite and finite volume. Using a quark flow analysis, we 
have obtained these results by generalizing the NLO PQ\chpt\ 
expressions calculated in the continuum in Ref.~\cite{BECIREVIC}. The 
main subtlety in applying this technique is due to the appearance of 
taste matrices inside the Feynman diagrams, since non-trivial signs can arise from 
the anticommutation relations of the taste generators.
We have shown that these signs can be accounted for
by a careful analysis of the relevant quark flow diagrams.

The \schpt\ expressions are generally necessary for 
performing chiral fits to lattice simulations where staggered light 
quarks are used. For simpler quantities than the form
factors, \schpt\  has been seen to be essential \cite{FPI04,Aubin:2005ar} 
in order to get reliable 
extrapolations both to the continuum limit and to the physical quark mass values.
For form factors,
the lattice data in Ref.~\cite{Aubin:2004ej} was not yet sufficiently
precise for the \schpt\ expressions to be required (over continuum forms)
for acceptable fits.  However, we expect that the forms derived here will become
more and more important  as the lattice data improves. 

Our results are valid to lowest order in HQET;  
in general, we neglect $1/m_B$ corrections.   
We do however include the $B^*$-$B$ splitting
$\Delstar$ on internal $B^*$ lines that are not in loops.
This prescription allows the  form factor $f_p$ to have
the physical $m_B^*$ pole structure.
Our treatment of the $B^*$-$B$ splitting is similar, but not identical, to that of
Refs.~\cite{FALK,BECIREVIC}. 
Unlike those authors, we iterate self-energy contributions, namely \figref{1loopP}(a) 
and the effect
of the one-loop mass shift of the $B$, to all orders.
This seems to us to be a natural choice, and also  makes the one-loop corrections
better behaved.  Indeed, with the values of light quark masses and momenta typically
used in staggered 
simulations \cite{SHIGEMITSU,Aubin:2004ej}, the one-loop $B$ mass shift can dominate
other one-loop corrections, so summing such self-energy contributions
to all orders seems entirely appropriate.  The final answers are then expressed
in terms of the splitting $\Delstar_{yx} \equiv M_{B^*_y} - M_{B_x} $.  In fitting
lattice data, we suggest using the actual lattice values of this mass difference
(at the simulated light quark mass values and lattice spacings), rather than
applying a one-loop formula for the mass shifts. 

Our primary results for the staggered,  partially quenched case with three
non-degenerate sea quarks are found in \secref{PQschpt}.
The form factor $f_v$ (also known as
$f_{\parallel}$) is given by \eq{fv} in terms of quantities
defined in \eqsthree{deltfv}{fv4a}{fv4b}, as well as the
wave function renormalization factors $\delta Z_{P_{xy}}$ and $\delta Z_{B_x}$ 
that are listed
in  \eqs{ZP}{ZB} of Appendix~\ref{app:wf_ren}.
Similarly, the form factor $f_p$ (also known as
$f_{\perp}$), is given by \eq{fp} in terms of quantities
defined in Eqs.~(\ref{eq:fpself}), (\ref{eq:fptreetilde}), (\ref{eq:deltfp}),
and (\ref{eq:fpD}) through (\ref{eq:fp5d}), as well as the
wave  function renormalization factors. We have also found a  single relation,
\eq{relation}, among the parameters
that control the analytic  valence mass dependence.  While this relation
is also satisfied by the parameters written down in Ref.~\cite{BECIREVIC}, it is
important to know that it persists even in the presence of the complete NLO forms
of the Lagrangian and current.

 Appropriate limits of our expressions can 
be taken for various relevant cases, including 
the case of full (unquenched) staggered QCD (in 
Sec.~\ref{sec:fullQCD}) 
and the case of continuum PQ\chpt\ 
with non-degenerate sea quark masses [\eqsfour{fvPQcont}{fpPQcont}{ZP-cont}{ZB-cont}].
Despite the fact that the latter are
continuum results, they have not, to our knowledge, appeared in the
literature before. Finally, our expressions can be corrected 
for finite volume effects using the results of Sec.~\ref{sec:FV}.

\bigskip
\bigskip
\centerline{\bf ACKNOWLEDGMENTS}
\bigskip
We thank J.\ Bailey, B.\ Grinstein, 
A.\ Kronfeld, P.\ Mackenzie, S.\ Sharpe and 
our colleagues in the MILC collaboration for 
helpful discussions.  We also are grateful to D.~Lin for
discussions on finite volume corrections and for sharing
with us the Mathematica code used to make the expansions in
Ref.~\cite{LIN_ARNDT}.
This work was partially supported by the
U.S. 
Department of Energy under grant numbers DE-FG02-91ER40628 and DE-FG02-92ER40699.

\appendix

\section{Feynman Rules}\label{app:rules}

In this appendix we list the \schpt\ propagators and (some of) the vertices in Minkowski space
\cite{HL_SCHPT}, as well as the corresponding  continuum versions.  

In \schpt,
the propagators for the heavy-light mesons are
\begin{eqnarray}
	\Bigl\{B_a B^\dagger_b\Bigr\}(k) &=& \frac{i\delta_{ab}}
	{2(\vk + i\epsilon)}\ , \label{eq:Bprop}\\
	\Bigl\{B^*_{\mu a} B^{*\dagger}_{\nu b}\Bigr\}(k) &=& 
	\frac{-i\delta_{ab}(g_{\mu\nu} - v_\mu v_\nu)}
	{2(\vk -\Delstar + i\epsilon)} \ \label{eq:Bstarprop}.
\end{eqnarray}
Here $a,b$ indicate the flavor-taste of the light quarks, and $\Delstar$ is
the $B^*$-$B$ splitting in the chiral limit, which we often neglect since
we work to leading order in HQET.

The $BB^*\pi$ vertex is:
\begin{equation}\label{eq:B-Bstar-pi}
\frac{g_\pi}{f}\,\left( B^\dagger_a \, B^{*}_{\mu b} - B^{*\dagger}_{\mu a}\, B_b \, 
\right) \, \partial^\mu \Phi_{ba} \ ,
\end{equation}
where repeated indices are summed.
Other needed vertices come from the expansion of
the LO current, \eq{LOcurrent}.  We have:
\begin{equation}\label{eq:current-vertex}
j^{\mu,c}_{\rm LO} = \kappa B^{*\mu}_{a}\left(\delta_{ac} -\frac{1}{8f^2} 
\Phi_{ab} \Phi_{bc} + \cdots \right) +
\kappa v^\mu B_a 
\left(\frac{1}{2f}\Phi_{ac} + \cdots \right) \ ,
\end{equation}
where  repeated indices are again summed and $\cdots$ represents terms involving higher numbers of pions, as well as contributions from the axial vector part of the current, which are not relevant
to the form factors.

If desired, each flavor-taste index can be replaced by a pair of indices representing flavor and taste separately.
We use Latin indices in the middle of the alphabet ($i,j,\dots$) as pure flavor indices, which take on the
values $1,2,\dots,N_{\rm sea}$ in full QCD.
Greek indices at the beginning of the alphabet ($\alpha,\beta,\gamma,\dots$) 
are used for quark taste indices,
running from $1$ to $4$.  Thus we can replace $a\to i\alpha$ and write, for example,
\begin{equation}
	\Bigl\{B_{i\alpha} B^\dagger_{j\beta}\Bigr\}(k) = \frac{i\delta_{ij}\delta_{\alpha\beta}}{2 (v\negcdot k + i\epsilon)}\ .
\end{equation}

As in Refs.~\cite{SCHPT,HL_SCHPT}, 
pion propagators are treated most easily by
dividing them into  connected and disconnected pieces, where the disconnected  parts come
from insertion (and iteration) of the hairpin vertices.
The connected propagators are 
\begin{equation}\label{eq:PropConnTaste}
	\Bigl\{\Phi^{\Xi}_{ij}\Phi^{\Xi'}_{j'i'}\Bigr\}_{\rm conn}(p) = 
	\frac{i\delta_{ii'}\delta_{jj'} \delta_{\Xi\Xi'}}{p^2 - m_{ij,\Xi}^2 + i\epsilon} \ ,
\end{equation}
where $\Xi$ is one of the 16 meson tastes [as defined after \eq{Phi}], 
and $m_{ij,\Xi}$ is
the tree-level mass of a taste-$\Xi$ meson composed of quarks of flavor $i$ and $j$:
\begin{equation}\label{eq:pi-masses-specific}
       m_{ij,\Xi}^2  = \mu (m_i + m_j) + a^2\Delta_\Xi.
\end{equation}
Here $\Delta_\Xi$ is the taste splitting, 
which can be expressed in terms of $C_1$, 
$C_3$, 
$C_4$ and $C_6$ 
in \eq{VSigma} \cite{SCHPT}. 
There is a residual $SO(4)$ taste
symmetry \cite{LEE_SHARPE} at this order, 
implying that the mesons within a given taste multiplet ($P$, 
$V$, 
$T$, 
$A$, 
or $I$)
are degenerate in mass.
We therefore usually use  the multiplet label to  represent the splittings.

Since the heavy-light propagators are most simply written with flavor-taste indices,
as in \eqs{Bprop}{Bstarprop}, 
it is convenient to rewrite \eq{PropConnTaste} in flavor-taste notation also:
\begin{equation}\label{eq:PropConn}
	\Bigl\{\Phi_{ab}\Phi_{b'a'}\Bigr\}_{\rm conn}(p) \equiv 
	\Bigl\{\Phi_{i\alpha, 
j\beta}\Phi_{j'\beta',i'\alpha'}\Bigr\}_{\rm conn}(p) = \sum_\Xi
	\frac{i\delta_{ii'}\delta_{jj'} T^\Xi_{\alpha\beta}  T^\Xi_{\beta'\alpha'} }
{p^2 - m_{ij,\Xi}^2 + i\epsilon} \ ,
\end{equation}
where $T^\Xi$ are the 16 taste generators, \eq{T_Xi}.

For flavor-charged pions ($i\not=j$), 
the complete propagators are just
the connected propagators in \eqsor{PropConnTaste}{PropConn}.  However, 
for flavor-neutral
pions ($i=j$), 
there are disconnected contributions coming from one or more
hairpin insertions.  At LO, 
these appear only for taste singlet, 
vector,
or axial-vector pions. 
Denoting the Minkowski hairpin
vertices as $-i\delta'_\Xi$, 
we have
\cite{SCHPT}:
\begin{equation}\label{eq:dp_def}
  \delta_\Xi' = \begin{cases}
    a^2 \delta'_V, &T_\Xi\in\{\xi_\mu\}\ \textrm{(taste\ vector);}\\*
    a^2 \delta'_A, &T_\Xi\in\{\xi_{\mu5}\}\ \textrm{(taste\ axial-vector);}\\*
    4m_0^2/3, &T_\Xi=\xi_{I}\ \textrm{(taste\ singlet);}\\*
    0, &T_\Xi\in\{\xi_{\mu\nu},\xi_5\}\ \textrm{(taste\ 
    tensor or pseudoscalar)}
  \end{cases}
\end{equation}
with
\begin{eqnarray}\label{eq:mix_vertex_VA}
 \delta'_{V(A)} & \equiv & \frac{16}{f^2} (C_{2V(A)} - C_{5V(A)})\ .
\end{eqnarray}
The disconnected pion propagator is then
\begin{equation}\label{eq:PropDiscTaste}
	\Bigl\{\Phi^{\Xi}_{ij}\Phi^{\Xi'}_{j'i'}\Bigr\}_{\rm disc}(p)=
	\delta_{ij}\delta_{j'i'} \delta_{\Xi\Xi'} \cD^\Xi_{ii,i'i'} \ ,
\end{equation}
where \cite{SCHPT}
 \begin{equation}\label{eq:Disc}
\cD^\Xi_{ii,i'i'} = -i\delta'_\Xi \frac{i}{(p^2-m_{ii,\Xi}^2)}
 \frac{i}{(p^2-m_{i'i',\Xi}^2)}
  \frac{(p^2-m_{U,\Xi}^2)(p^2-m_{D,\Xi}^2)(p^2-m_{S,\Xi}^2)}
       {(p^2-m_{\pi^0,\Xi}^2)(p^2-m_{\eta,\Xi}^2)(p^2-m_{\eta',\Xi}^2)}\ .
\end{equation}
For concreteness we have assumed that there are three sea-quark flavors: $u$, 
$d$, 
and $s$;
the generalization to $N_{\rm sea}$ flavors is immediate.  
Here $m_{U,\Xi}\equiv m_{uu,\Xi}$ is the mass of a taste-$\Xi$  pion made from a $u$ and a $\bar u$ quark,
neglecting hairpin mixing (and similarly for $m_{D,\Xi}$ and $m_{S,\Xi}$), 
 $m_{\pi^0,\Xi}$, 
$m_{\eta,\Xi}$, 
and $m_{\eta',\Xi}$
are the mass eigenvalues after mixing is included, 
and
the $i\epsilon$ terms have been left implicit.  When specifying the particular member
of a taste multiplet appearing in the disconnected
propagator is unnecessary, 
we abuse this notation slightly following
\eq{pi-masses-specific} and
refer to $\cD^V_{ii,i'i'} $, 
$\cD^A_{ii,i'i'}$, 
or $\cD^I_{ii,i'i'}$. 
In flavor-taste notation we have:
\begin{equation}\label{eq:PropDisc}
	\Bigl\{\Phi_{ab}\Phi_{b'a'}\Bigr\}_{\rm disc}(p) \equiv 
	\Bigl\{\Phi_{i\alpha, 
j\beta}\Phi_{j'\beta',i'\alpha'}\Bigr\}_{\rm disc}(p) = 
	\delta_{ij}\delta_{j'i'} 
\sum_\Xi
T^\Xi_{\alpha\beta}  T^\Xi_{\beta'\alpha'} \cD^\Xi_{ii,i'i'}
\end{equation}

For comparison, we now describe the continuum versions of the Feynman rules
\cite{MAN_WISE}.
Since taste violations do not appear in $\cL_{\rm HL}$, \eq{L-HL},
the continuum-theory  version of 
\eqs{Bprop}{Bstarprop} are unchanged
except that flavor-taste indices are replaced by pure flavor indices ($i,j$):
\begin{eqnarray}
	\Bigl\{B_i B^\dagger_j\Bigr\}(k) &=& \frac{i\delta_{ij}}
	{2(\vk + i\epsilon)} \qquad[{\rm continuum}], \label{eq:Bprop-cont}\\
	\Bigl\{B^*_{\mu i} B^{*\dagger}_{\nu j}\Bigr\}(k) &=& 
	\frac{-i\delta_{ij}(g_{\mu\nu} - v_\mu v_\nu)}
	{2(\vk + i\epsilon)} \qquad[{\rm continuum}].\label{eq:Bstarprop-cont}
\end{eqnarray}

Similarly, the continuum $BB^*\pi$ \cite{MAN_WISE} and current 
vertices are identical to those 
in \schpt, 
aside from the redefinition of the indices and a factor of $2$ for each $\Phi_{ab}$
field due to the non-standard normalization of the 
generators in the \schpt\ case, \eq{T_Xi}.   
The continuum version of  
\eq{B-Bstar-pi} is
\begin{equation}\label{eq:B-Bstar-pi-cont}
2\frac{ig_\pi}{f}\,\left(B^{*\dagger}_{\mu i}\, B_j \, -
B^\dagger_i \, B^{*}_{\mu j}\right) \, \partial^\mu \Phi_{ji} \qquad[{\rm continuum}];
\end{equation}
while the continuum version of 
\eq{current-vertex} is
\begin{equation}\label{eq:current-vertex-cont}
j^{\mu,k}_{\rm LO} = \kappa B^{*\mu}_{\ell}\left(\delta_{\ell k} -\frac{1}{2f^2} 
\Phi_{\ell i } \Phi_{i k} + \cdots \right) +
\kappa v^\mu B_\ell 
\left(\frac{1}{f}\Phi_{\ell k} + \cdots \right) 
\qquad[{\rm continuum}].
\end{equation}

Because of taste-violations in the \schpt\ pion sector, the differences between
the propagators  \eqsthree{PropConnTaste}{PropDiscTaste}{Disc} and their continuum versions are slightly
less trivial.  The continuum connected propagator is 
\begin{equation}\label{eq:PropConnContinuum}
	\Bigl\{\Phi_{ij}\Phi_{j'i'}\Bigr\}_{\rm conn}(p) = 
	\frac{i\delta_{ii'}\delta_{jj'} }{p^2 - m_{ij}^2 + i\epsilon} \qquad[{\rm continuum}],
\end{equation}
with 
\begin{equation}\label{eq:pi-masses-continuum}
       m_{ij}^2  = \mu (m_i + m_j) \qquad[{\rm continuum}].
\end{equation}
The continuum disconnected propagator is 
\begin{equation}\label{eq:PropDiscContinuum}
	\Bigl\{\Phi_{ij}\Phi_{j'i'}\Bigr\}_{\rm disc}(p)=
	\delta_{ij}\delta_{j'i'} \cD_{ii,i'i'} \qquad[{\rm continuum}],
\end{equation}
where \cite{SCHPT}
 \begin{equation}\label{eq:DiscContinuum}
\cD_{ii,i'i'} = -i\delta' \frac{i}{(p^2-m_{ii}^2)}
 \frac{i}{(p^2-m_{i'i'}^2)}
  \frac{(p^2-m_{U}^2)(p^2-m_{D}^2)(p^2-m_{S}^2)}
       {(p^2-m_{\pi^0}^2)(p^2-m_{\eta}^2)(p^2-m_{\eta'}^2)}\qquad[{\rm continuum}],
\end{equation}
with now $\delta' = m_0^2/3$.

Note the difference in normalization between $\delta'$ and the \schpt\ taste-singlet
hairpin, $\delta'_I$, \eq{dp_def}.  This arises from the fact that $m_0^2/3$ is defined to
be the strength of the hairpin vertex when one has a single species of quark on 
each side of the vertex \cite{CHIRAL_FSB}.
In the staggered case, each normalized taste-singlet field is made out of four species (tastes), for example
$\phi^I = \frac{1}{2}(\phi_{11}+\phi_{22}+\phi_{33}+\phi_{44})$, where $\phi$ is flavor neutral,
and only taste indices are shown. In the disconnected propagator of two such fields,
there are 16 terms, and a factor of $(1/2)^2$ from the normalization, so there is an overall factor
of 4 relative to a single-species disconnected propagator, such as that of $\phi_{11}$ with $\phi_{22}$.  
At one loop, the ``external'' fields in this propagator are always valence fields, so the 
normalization issue has nothing directly to do with the fourth root trick for 
staggered sea quarks. (The normalization is in fact compensated by the extra factors of 2 in the continuum
vertices.)  
The rooting does however affect the $\eta'_I$ mass that appears in denominator
of \eq{DiscContinuum}, which comes from iterations of the hairpin and therefore involves sea quarks. The end result
is that $m^2_{\eta',I}\approx N_{\rm sea}m_0^2/3$ (for large $m_0$), rather than 
$\approx4N_{\rm sea}m_0^2/3$,
the value 
in the unrooted theory \cite{SCHPT}.  
In the continuum, we also
have $m^2_{\eta'}\approx N_{\rm sea}m_0^2/3$.

\section{Integrals}\label{app:int}

Here we collect the
integrals needed in evaluating the diagrams for the semileptonic form factors
\cite{HL_SCHPT,BECIREVIC}.

The disconnected propagators can be written as a sum of single
or double poles using the (Euclidean) residue functions introduced in Ref.~\cite{SCHPT}
or their Minkowski-space versions.
We define $\{m\}\equiv \{m_1,m_2,\dots,m_n\}$ as the set of masses that appear in the
denominator of \eq{Disc}, 
and $\{\mu\}\equiv \{\mu_1,\mu_2,\dots,\mu_k\}$ as the numerator
set of masses.  Then, 
for $n>k$ and all masses distinct, 
we have:
\begin{equation}\label{eq:lagrange}
        \cI^{[n,k]}\left(\left\{m\right\}\!;\!
        \left\{\mu\right\}\right)
        \equiv \frac{\prod_{i=1}^k (q^2 - \mu^2_i)}
               {\prod_{j=1}^n (q^2 - m^2_j + i\epsilon)} =
        \sum_{j=1}^n \frac{
        \hat R_j^{[n,k]}\left(\left\{m\right\}\!;\!
          \left\{\mu\right\}\right)}{q^2 - m^2_j + i\epsilon}\ ,
\end{equation}
where the Minkowski space residues $\hat R_j^{[n,k]}$ are given by
\begin{equation}\label{eq:Mink-residues}
        \hat R_j^{[n,k]}\left(\left\{m\right\}\!;\!\left\{\mu\right\}\right)
         \equiv  \frac{\prod_{i=1}^k ( m^2_j -\mu^2_i)}
        {\prod_{r\not=j} ( m^2_j - m^2_r )}\ .
\end{equation}
If there is one double pole term for $q^2=m_\ell^2$ (where $m_\ell \in \{m\}$), 
then
\begin{eqnarray}
        \cI^{[n,k]}_{\rm dp}\left(m_{\ell};\left\{m\right\}\!
        ;\!\left\{\mu\right\}\right) 
        &\equiv& \frac{\prod_{i=1}^k (q^2 - \mu^2_i)}
        {(q^2 - m^2_{\ell}+i\epsilon )\prod_{j=1}^{n} (q^2 - 
        m^2_j+i\epsilon )}
        \nonumber\\*
        & = & \frac{\partial}{\partial m^2_\ell}   \sum_{j=1}^n 
        \frac{\hat R_j^{[n,k]}\left(\left\{m\right\}\!;\!
          \left\{\mu\right\}\right)}{q^2 - m^2_j +i\epsilon}\label{eq:lagrange2}\ .
\end{eqnarray}

In the end we want to write the results in terms of the Euclidean-space residues $R_j^{[n,k]}$,
because they are ones we have used previously \cite{SCHPT,HL_SCHPT}. In Euclidean
space the sign of each factor in \eq{Mink-residues} is changed.  We therefore have
\begin{equation}\label{eq:residues}
 R_j^{[n,k]}\left(\left\{m\right\}\!;\!\left\{\mu\right\}\right)
         \equiv  \frac{\prod_{i=1}^k (\mu^2_i- m^2_j)}
        {\prod_{r\not=j} (m^2_r - m^2_j)} = (-1)^{n+k-1} 
\hat  R_j^{[n,k]}\left(\left\{m\right\}\!;\!\left\{\mu\right\}\right) \ .
\end{equation}

The integrals needed for the form factors are (\cite{BOYD,BECIREVIC})
\begin{eqnarray}
	\cI_1 & = & \mu^{4-d}\!\int\frac{d^d q}{(2\pi)^d}\;
	\frac{i}{q^2-m^2+i\epsilon} \to
	\frac{1}{(4\pi)^2}I_1(m) \label{eq:I1int}\ , \\
	\cI_2 & = & \mu^{4-d}\!\int\frac{d^d q}{(2\pi)^d}\;
	\frac{i}{(\vq - \Delta+i\epsilon)(q^2-m^2+i\epsilon)} \to 
	\frac{1}{(4\pi)^2}\frac{1}{\Delta}I_2(m,\Delta) \ , 
	\label{eq:I2int}\\
	\cI_3^\mu & = & \mu^{4-d}\!\int\frac{d^d q}{(2\pi)^d}\;
	\frac{iq^\mu}{(\vq - \Delta+i\epsilon)(q^2-m^2+i\epsilon)} \to 
	\frac{v^\mu}{(4\pi)^2}
	\left[I_2(m,\Delta) + I_1(m)\right]\ , 
	\label{eq:I1I2int}\\
	\cJ^{\mu\nu} & = & \mu^{4-d}\!\int\frac{d^d q}{(2\pi)^d}\;
	\frac{iq^\mu q^\nu}
	{(\vq - \Delta+i\epsilon)(q^2-m^2+i\epsilon)} \to 
	\frac{\Delta}{(4\pi)^2}
	\left[J_1(m,\Delta) g^{\mu\nu} 
	+ J_2(m,\Delta)v^\mu v^\nu\right]\,,\nonumber \\
&&\label{eq:J1J2int}
\end{eqnarray}
where the arrows represent the fact that the r.h.s.\ of these expressions have already been renormalized (unlike the corresponding equations in Ref.~\cite{BECIREVIC}).

\section{Wavefunction Renormalization Factors}
\label{app:wf_ren}

The one loop chiral
corrections to the wave function renormalization factors $Z_{B}$ and $Z_P$ are 
are \cite{SCHPT,HL_SCHPT}
\begin{eqnarray}\eqn{ZP}
  \delta Z_{P_{xy}} & =&\frac{1}{3(4\pi f)^2}\Biggl\{
  \frac{1}{16}\sum_{f,\Xi} 
  \left[
  I_1\left(m_{xf,\Xi}\right)
  +I_1\left(m_{yf,\Xi}\right)
  \right]
  \nonumber\\&&{}
  +\frac{1}{3}\Biggl[\sum_{j\in\cM^{(3,x)}}
    \frac{\partial}{\partial m_{X,I}^2}
  \left(
  R^{[3,3]}_{j} \left(\cM^{(3,x)}_I ; \mu^{(3)}_I\right)
  I_1(m_{j,I})\right)\nonumber \\* &&
  +\sum_{j\in\cM^{(3,y)}}
    \frac{\partial}{\partial m_{Y,I}^2}
    \left(
    R^{[3,3]}_{j}  \left(\cM^{(3,y)}_I ; \mu^{(3)}_I\right)
    I_1(m_{j,I})\right)\nonumber\\&&{}
  +2\sum_{j\in\cM^{(4,xy)}}
  R^{[4,3]}_{j} \left(\cM^{(4,xy)}_I ; \mu^{(3)}_I\right)
  I_1(m_{j,I})\Biggr]
  \nonumber \\* &&{}
+a^2 \delta'_V
  \Biggl[
    \sum_{j\in\cM^{(4,x)}}
    \frac{\partial}{\partial m_{X,V}^2}
  \left(
  R^{[4,3]}_{j}  \left(\cM^{(4,x)}_V ; \mu^{(3)}_V\right)
  I_1(m_{j,V})
  \right)\nonumber \\* &&
  +\sum_{j\in\cM^{(4,y)}}
    \frac{\partial}{\partial m_{Y,V}^2}
    \left(
    R^{[4,3]}_{j}\left(\cM^{(4,y)}_V ; \mu^{(3)}_V\right) 
    I_1(m_{j,V})\right)\nonumber\\&&{}
  -2\sum_{j\in\cM^{(5,xy)}}
  R^{[5,3]}_{j}\left(\cM^{(5,xy)}_V ; \mu^{(3)}_V\right)
  I_1(m_{j,V})
    \Biggr]\nonumber \\* &&
  + \Bigl[ V \to A \Bigr]\Biggr\}, \
\end{eqnarray}

\begin{eqnarray}\eqn{ZB}
  \delta Z_{B_x}
  &= &  \frac{-3g_\pi^2}{(4\pi f)^2}
  \biggl\{\frac{1}{16}\sum_{f,\Xi} I_1(m_{xf,\Xi})
      \nonumber \\*&&{} +
    \frac{1}{3}\sum_{j\in\cM^{(3,x)}}
    \frac{\partial}{\partial m_{X,I}^2}\left[
    R^{[3,3]}_{j}  \left(\cM^{(3,x)}_I ; \mu^{(3)}_I\right)
    I_1(m_{j,I})
     \right]
    \nonumber \\*&&{} 
     +   a^2\delta'_V\sum_{j\in\cM^{(4,x)}}
    \frac{\partial}{\partial m_{X,V}^2}
    \left[ R^{[4,3]}_{j}\left(\cM^{(4,x)}_V ; \mu^{(3)}_V\right)
    I_1(m_{j,V})\right]
 	+ [V\to A]  
   \biggr\}    \ ,
\end{eqnarray}
where $f$ runs over the sea quarks ($u$, $d$, $s$).

For the continuum result in partially quenched \chpt, we can
simply set $a=0$ and ignore taste splittings. In the \opopo\ case, we get
\begin{eqnarray}\eqn{ZP-cont}
  \delta Z^{\rm cont}_{P_{xy}} & =&\frac{1}{3(4\pi f)^2}\Biggl\{
  \sum_{f} 
  \left[
  I_1\left(m_{xf}\right)
  +I_1\left(m_{yf}\right)
  \right]
  \nonumber\\&&{}
   +\frac{1}{3}\Biggl[\sum_{j\in\cM^{(3,x)}}
    \frac{\partial}{\partial m_{X}^2}
  \left(
  R^{[3,3]}_{j} \left(\cM^{(3,x)} ; \mu^{(3)}\right)
  I_1(m_{j})\right)\nonumber \\* &&
  +\sum_{j\in\cM^{(3,y)}}
    \frac{\partial}{\partial m_{Y}^2}
    \left(
    R^{[3,3]}_{j}  \left(\cM^{(3,y)} ; \mu^{(3)}\right)
    I_1(m_{j})\right)\nonumber\\&&{}
  +2\sum_{j\in\cM^{(4,xy)}}
  R^{[4,3]}_{j} \left(\cM^{(4,xy)} ; \mu^{(3)}\right)
  I_1(m_{j})\Biggr]
  \Biggr\}, \
\end{eqnarray}

\begin{eqnarray}\eqn{ZB-cont}
  \delta Z^{\rm cont}_{B_x}
  &= &  \frac{-3g_\pi^2}{(4\pi f)^2}
  \biggl\{\sum_{f} I_1(m_{xf})
      \nonumber \\*&&{} +
    \frac{1}{3}\sum_{j\in\cM^{(3,x)}}
    \frac{\partial}{\partial m_{X}^2}\left[
    R^{[3,3]}_{j}  \left(\cM^{(3,x)} ; \mu^{(3)}\right)
    I_1(m_{j})
     \right]
   \biggr\}    
\end{eqnarray}

Returning to $a\not=0$,
and taking the valence quark masses to be $m_x=m_y=m_u=m_d$, 
we have the $2\!+\!1$ full QCD pion result in \schpt:
\begin{eqnarray}
  \delta Z_{\pi} & =&
  \frac{1}{3(4\pi f)^2}\Biggl\{
  \frac{1}{16}\sum_{\Xi} 
  \left[ 4 I_1(m_{\pi,\Xi}) +  2I_1(m_{K,\Xi}) \right]
    \nonumber \\* &&\hspace{-.2truecm}
  +(-4a^2 \delta'_V)\Biggl[
    \frac{(m^2_{S_V} - m^2_{\pi_V})}
       {(m^2_{\eta_V} - m^2_{\pi_V})(m^2_{\eta'_V} - m^2_{\pi_V})} 
  I_1(m_{\pi_V})+
    \frac{(m^2_{S_V} - m^2_{\eta_V})}
       {(m^2_{\pi_V} - m^2_{\eta_V})(m^2_{\eta'_V} - m^2_{\eta_V})} 
   I_1(m_{\eta_V}) \nonumber \\* &&  +
    \frac{(m^2_{S_V} - m^2_{\eta'_V})}
       {(m^2_{\pi_V} - m^2_{\eta'_V})(m^2_{\eta_V} - m^2_{\eta'_V})} 
   I_1(m_{\eta'_V})\Biggr]
  + \Bigl[ V \to A \Bigr]\Biggr\} \ .
\end{eqnarray}
Taking the valence quark masses to be $m_x=m_u=m_d$ and $m_y=m_s$ gives 
the $2\!+\!1$ full QCD kaon result:
\begin{eqnarray}
  \delta Z_{K} & =&\frac{1}{3(4\pi f)^2}\Biggl\{
  \frac{1}{16}\sum_{\Xi}\left(
  2I_1(m_{\pi,\Xi})+ 3I_1(m_{K,\Xi}) +  I_1(m_{S,\Xi})\right)
  \nonumber \\* &&
  -\frac{1}{2}I_1(m_{\pi_I}) 
  +\frac{3}{2}I_1(m_{\eta_I}) -I_1(m_{S_I})
  \nonumber \\* &&
  +(- a^2\delta'_V)\Biggl(  
  \frac{(m^2_{S_V}+m^2_{\pi_V}-2m^2_{\eta_V})^2}
       {(m^2_{\pi_V}-m^2_{\eta_V})
	 (m^2_{S_V}-m^2_{\eta_V})(m^2_{\eta'_V}-m^2_{\eta_V})}
	 I_1(m_{\eta_V}) \nonumber\\*&&
         +\frac{(m^2_{S_V}+m^2_{\pi_V}-2m^2_{\eta'_V})^2}
	 {(m^2_{\pi_V}-m^2_{\eta'_V})
	   (m^2_{S_V}-m^2_{\eta'_V})(m^2_{\eta_V}-m^2_{\eta'_V})}
	 I_1(m_{\eta'_V}) \nonumber\\*&&
	 +  \frac{m^2_{S_V}-m^2_{\pi_V}}
	 {(m^2_{\eta_V}-m^2_{\pi_V})(m^2_{\eta'_V}-m^2_{\pi_V})}
	 I_1(m_{\pi_V}) 
	 +  \frac{m^2_{\pi_V}-m^2_{S_V}}
	 {(m^2_{\eta_V}-m^2_{S_V})(m^2_{\eta'_V}-m^2_{S_V})}
	 I_1(m_{S_V}) \Biggr)  \nonumber \\* &&
	 + \Bigl( V\to A  \Bigr)\Biggr\}\ .
\end{eqnarray}
  
Setting $m_x=m_u=m_d$ in \eq{ZB} results in 
the $2\!+\!1$ full QCD result for the $B$ wavefunction renormalization:
\begin{eqnarray}
  \delta Z_{B}
  &= &  \frac{3g_\pi^2}{(4\pi f)^2}
  \Biggl\{-\frac{1}{16}\sum_{\Xi} \left[ 2 I_1(m_{\pi,\Xi}) +
  I_1(m_{K,\Xi}) \right]
  + \frac{1}{2}
  I_1(m_{\pi_I})
  -  \frac{1}{6} I_1(m_{\eta_I})
    \nonumber \\*&&{} 
    +   a^2\delta'_V 
    \Biggl[
      \frac{(m^2_{S_V} - m^2_{\pi_V})}{(m^2_{\eta_V} -
	m^2_{\pi_V})(m^2_{\eta'_V} - m^2_{\pi_V})} 
      I_1(m_{\pi_V})+
      \frac{(m^2_{S_V} - m^2_{\eta_V})}
	   {(m^2_{\pi_V} - m^2_{\eta_V})(m^2_{\eta'_V} - m^2_{\eta_V})} 
	   I_1(m_{\eta_V}) \nonumber \\* &&  +
	   \frac{(m^2_{S_V} - m^2_{\eta'_V})}
		{(m^2_{\pi_V} - m^2_{\eta'_V})(m^2_{\eta_V} - m^2_{\eta'_V})} 
		I_1(m_{\eta'_V})\Biggr]
    + [V\to A]  
    \Biggr\} \ .   
\end{eqnarray}
Finally, 
putting $m_x=m_s$ and $m_u=m_d$ in \eq{ZB}, we obtain
the full QCD $B_s$ renormalization factor in the $2\!+\!1$ case:
\begin{eqnarray}
  \delta Z_{B_s}
  &= &  \frac{3g_\pi^2}{(4\pi f)^2}
  \Biggl\{-\frac{1}{16}\sum_{\Xi} \left[ I_1(m_{S,\Xi})
    + 2I_1(m_{K,\Xi}) \right]+
  I_1(m_{S_I}) - \frac{2}{3}I_1(m_{\eta_I})  
  \nonumber \\*&&{}
\hspace{-0.4cm}  +(-a^2\delta'_V) \biggl[
    \frac{(m^2_{S_V} - m^2_{\pi_V})}{(m^2_{S_V} - 
      m^2_{\eta_V})(m^2_{S_V} - m^2_{\eta'_V})}
    I_1(m_{S_V})
    + \frac{(m^2_{\eta_V} - m^2_{\pi_V})}{(m^2_{\eta_V} -
      m^2_{S_V})(m^2_{\eta_V} - m^2_{\eta'_V})}
    I_1(m_{\eta_V}) \nonumber\\*&&{}+ 
    \frac{(m^2_{\eta'_V} - m^2_{\pi_V})}{(m^2_{\eta'_V} 
      - m^2_{S_V})(m^2_{\eta_V'} - m^2_{\eta_V})}
    I_1(m_{\eta'_V})\biggr]
  + [V\to A] \Biggr\} \ .   
\end{eqnarray}

\vfill\eject

\begin{table}[htdp]
\caption{Connecting the one-loop diagrams from Ref.~\cite{BECIREVIC} (left column) and this paper (right column).}
\begin{center}
\begin{tabular}{|c|c|}
\hline
Ref.~\cite{BECIREVIC} & This work\\
\hline
(4) & \figref{1loopV}(a) \\\hline
(7) & \figref{1loopP}(a) \\\hline
(9) & \figref{1loopP}(b) \\\hline
(12) &  \figref{1loopP}(c) \\\hline
(13) &  \figref{1loopP}(d) \\\hline
(14) & \figref{1loopV}(b) \\\hline
\end{tabular}
\end{center}
\label{tab:bec_us}
\end{table}

\end{document}